\documentclass[12pt]{article}
\usepackage{amsfonts}
\usepackage{amssymb}
\usepackage{amsmath}
\usepackage[numbers,sort&compress]{natbib}
\usepackage[colorlinks,linkcolor=blue,citecolor=blue,urlcolor=blue]{hyperref}
\usepackage{graphicx,indentfirst,tabularx}

\newtheorem{theorem}{Theorem}

\newtheorem{algorithm}[theorem]{Algorithm}

\newtheorem{corollary}[theorem]{Corollary}

\newtheorem{remark}[theorem]{Remark}

\newenvironment{proof}[1][Proof]{\noindent\textbf{#1.} }{\ \rule{0.5em}{0.5em}}
\input{tcilatex}
\begin{document}

\title{Zeros and Extrema of Functions via Random Measures}
\author{Athanasios C. Micheas \\
%EndAName
Department of Statistics\\
University of Missouri\\
email: micheasa@missouri.edu}
\maketitle

\begin{abstract}
We present methods that provide all zeros and extrema of a function that do
not require differentiation. Using point process theory, we are able to
describe the locations of zeros or maxima, their number, as well as their
distribution over a given window of observation. The algorithms in order to
accomplish the theoretical development are also provided, and they are
exemplified using many illustrative examples, for real and complex functions.
\end{abstract}



\section{Introduction}

Perhaps one of the most fundamental problems in mathematics is finding when
the function assumes the value zero or where it achieves its extreme
(maximum and minimum) values. One of the most important approaches, Newton's
method, withstood the test of time, and is still employed as we speak, in
every mathematical discipline. Over the past four centuries, scientists have
made many improvements to the original method, including introducing
stochasticity. Although it has its difficulties and limitations, it has been
the standard method to employ, since it involves one of the easiest and well
understood concepts, that of the tangent via a Taylor expansion.

In this paper, we propose a novel method that provides all zeros or the
extrema of a function and their number, potentially in a single iteration,
that does not suffer the usual problems of existing methods in the
literature. There is a vast collection of papers and textbooks on general
optimization methods, with some recent published works including \cite%
{pedregal2004introduction}, \cite{pierre2012optimization}, \cite%
{xiao2014multi}, \cite{bertsekas2015convex}, \cite{zhou2015hybrid}, \cite%
{kochenderfer2019algorithms}, \cite{diwekar2020introduction}, \cite%
{benfenati2022binary}, \cite{gao2025new}, \cite{sadana2025survey} and the
references therein.

To formulate the ideas mathematically, let $\mathbb{M}$ denote the real
numbers $\mathbb{R}$ or the complex plane $\mathbb{C}$, and let $\mathcal{B}(%
\mathbb{X})$ denote the Borel sets of a space $\mathbb{X},$ using the
topology induced by an appropriate metric, depending on $\mathbb{X}.$ In
what follows, we consider a general function space%
\begin{equation}
\mathcal{F}_{\mathbb{M}}^{\mathbb{M}^{p}}=\{f:\mathbb{M}^{p}\rightarrow
\mathbb{M},\text{ }f\text{ continuous in }(\mathbb{M},\mathcal{B}(\mathbb{M}%
))\},  \label{FuncSpace}
\end{equation}%
where we will assume that $\mathcal{F}_{\mathbb{M}}^{\mathbb{M}^{p}}$ is
Hilbert, based on some inner product $\rho .$

Now since we do not know the locations in the set of zero points of $f$, nor
do we know how many of them there are, we can think of this set as the
realization of a random set that requires estimation or approximation.
Specifically, the set of zeros of $f$ can be described by the deterministic
set%
\begin{equation}
\Xi _{f}=\{\mathbf{\xi }\in \mathcal{X}\subset \mathbb{M}^{p}:f(\mathbf{\xi }%
)=0\},  \label{zerosRandomSet}
\end{equation}%
which has unknown elements and unknown cardinality $\#(\Xi _{f}).$ In order
to describe the stochastic process that yields realizations of elements from
$\Xi _{f},$ we naturally turn to point process (PP) theory. A point process
is a random collection of points from a space $\mathcal{X}$, which is also
random in number, and therefore, it offers an appealing framework to help us
build the theory required to approximate the set $\Xi _{f}$.

We call the new method Point Process zeros (PPZ). The proposed approach
utilizes point process theory and has the following characteristics:

\begin{itemize}
\item It is intuitively appealing, enjoys great interpretation and is easy
to use.

\item The PPZ provides all unknown zeros of a given function and their
number, at the same time, over a bounded window of observation, and in a
single realization of the point process.

\item The PPZ provides all unknown points of extrema (points where the
derivative is zero) and the function maximum/minimum value, at the same
time, based on a single realization of the point process over a bounded
window of observation.

\item Existing methods attempt to approach iteratively a single zero or
point of extrema, and are often locked into local characteristics of the
target function (e.g., simulated annealing) or do not converge. The PPZ does
not suffer any of these difficulties.

\item The PPZ does not require and it is not affected by starting values.

\item The methods can be employed in any dimensional space, since the theory
scales easily to higher dimensions by construction. However, the curse of
dimensionality does introduce longer computational times.

\item The methods do not require differentiation of the given function to
find its zeros or extrema, but if the gradient is known it can be used to
speed up calculations.

\item The function under consideration must satisfy only mild conditions,
such as continuity and being uniformly bounded over the window of
observation.
\end{itemize}

For foundations, modeling, applications, computation methods and evaluation
of point process models we refer to the texts by \cite{BarndorffNielsen},
\cite{vanLieshout}, \cite{lantuejoul2001geostatistical}, \cite{LawsonDenison}%
, \cite{Moller2003}, \cite{MollerWaag2004}, \cite{DaleyVereJones2005}, \cite%
{DaleyVereJones2008}, \cite{Illian2008}, \cite{Gelfand2010}, \cite%
{Chiuetal2013}, \cite{Spodarev2013}, \cite{Diggle2013} and \cite%
{baddeley2016spatial}. Recent papers exploring general methodologies,
applications and simulations of such processes include \cite%
{guan2010sufficient}, \cite{yau2012generalization}, \cite{perry2013point},
\cite{Micheas2014}, \cite{zhou2015spatio}, \cite{lavancier2015determinantal}%
, \cite{cronie2018non}, \cite{micheas2018sppmix}, \cite{Micheas2019}, \cite%
{zhuang2020detection}, \cite{ChenMicheasHolan2020}, \cite{TangLi2021}, \cite%
{Baddeley2022}, \cite{brochard2022particle}, \cite{WuMicheas2022}, \cite%
{chen2022hierarchical}, \cite{Baetal2023}, \cite{KresinSchoenberg2023}, \cite%
{WuMicheas2024}, \cite{VanLieshout2024}, \cite{cronie2024cross}, \cite%
{kindap2024point}, \cite{gonzalez2025nonparametric}, \cite{Micheas2025},
\cite{wang2026inference}, and the references therein.

The paper proceeds as follows; in Section 2 we introduce the new method for
finding zeros and study its theoretical properties with respect to the
number of zeros and their locations. Generalizations to multidimensional and
vector valued functions are also presented. Implementation considerations
are presented in Section 3, including the main PPZ algorithm and its
adaptive version that allows us to get as close to a zero or point of
extrema as we wish. Extensive simulations and illustrative examples for real
and complex functions are presented in Section 4. Concluding remarks are
given in the last section.

\section{Zeros of Functions via Random Measures}

Consider a probability space $(\Omega ,\mathcal{A},P),$ and let $%
%TCIMACRO{\U{2115} }%
%BeginExpansion
\mathbb{N}
%EndExpansion
^{lf}$ denote the collection of all locally finite counting measures on $%
\mathcal{X}\subset \mathbb{M}^{p},$ where if $\varphi \in
%TCIMACRO{\U{2115} }%
%BeginExpansion
\mathbb{N}
%EndExpansion
^{lf}$ and $\mathcal{W}\in \mathcal{B}(\mathcal{X}),$ then $\varphi (%
\mathcal{W})$ denotes the number of points in the set $\mathcal{W},$ also
known as the window of observation. In order to define a random counting
measure we create a measurable space based on a $\sigma $-field of subsets
of $\mathbb{N}$.

In particular, consider $\mathcal{N}$, the generated $\sigma $-field from
the collection of sets of the form $\{\varphi \in
%TCIMACRO{\U{2115} }%
%BeginExpansion
\mathbb{N}
%EndExpansion
^{lf}:\varphi (\mathcal{W})=n\}$, for all $\mathcal{W}\in \mathcal{B}(%
\mathcal{X})$ and all $n\in \mathbb{N}=\{0,1,2,...\},$ so that $(%
%TCIMACRO{\U{2115} }%
%BeginExpansion
\mathbb{N}
%EndExpansion
^{lf},\mathcal{N}^{lf})$ is a measurable space. Then a mapping $N$ from $%
(\Omega ,\mathcal{A},P)$ into $(%
%TCIMACRO{\U{2115} }%
%BeginExpansion
\mathbb{N}
%EndExpansion
^{lf},\mathcal{N}^{lf})$ can be thought of as a random object $N(\omega
)(.)=\varphi (.),$ that yields counting measures $\varphi $ for different $%
\omega \in \Omega $ and hence it is a random counting measure. The induced
probability measure $\Pi _{N}(Y)=P(N\in Y)=P(N^{-1}(Y)),$ describes the
probability distribution of $N$ and can be used to compute probabilities
over sets $Y\in \mathcal{N}^{lf},$ where each element of $Y$ is a counting
measure.

In order to appreciate the usefulness of this construction, consider a
countable set of events $S=\{\mathbf{x}_{1},\mathbf{x}_{2},\dots \},$ $%
\mathbf{x}_{i}\in \mathcal{X}.$ Knowing $\varphi (\mathcal{W})$ for all $%
\mathcal{W}\in \mathcal{B}(\mathcal{X})$ is equivalent to knowing the
locations of all points in $S$. Indeed, $\varphi (\mathcal{W}%
)=\tsum\limits_{i=1}^{+\infty }I(\mathbf{x}_{i}\in \mathcal{W}),$ $\forall
\mathcal{W}\in \mathcal{B}(\mathcal{X}),$ so that knowing the locations
implies that we know the counting measure $\varphi .$ Conversely, $\mathbf{x}
$ is a point from $S$ if $\varphi (\{\mathbf{x}\})>0$ and the point is
distinct if $\varphi (\{\mathbf{x}\})=1.$ Now if $\varphi $ is a realization
of a random counting measure, i.e., $\varphi =N(\omega ),$ for some $\omega
\in \Omega ,$ then this important relationship between $\varphi $ and $S$
defines a random collection of points (a countable random set) $N,$ known as
a point process.

We will exploit this representation of collections of points from the space $%
\mathcal{X}$ in order to find zeros or extrema of a function $f\in \mathcal{F%
}_{\mathbb{M}}^{\mathbb{M}^{p}}.$ Specifically, the set of zeros of $f$ can
be described by the set $\Xi _{f}$ which is possibly uncountable for $%
\mathcal{X}=\mathbb{M}^{p}$, but will be assumed to be countable over $%
\mathcal{X}\subset \mathbb{M}^{p}.$ Therefore, the cardinality $\#(\Xi _{f})$
of $\Xi _{f}$ denotes the number of zeros of $f$ in $\mathcal{X}$, and could
be any natural integer in $\mathbb{N}$.

Since we do not know either the number of zeros $\#(\Xi _{f})$ nor their
locations, the countable set $\Xi _{f}$ will be treated as a realization of
a random counting measure $N$, such that $N$ has counting variables $N(%
\mathcal{W})=\#(N\cap \mathcal{W}),$ for any $\mathcal{W}\in \mathcal{B}(%
\mathcal{X}),$ i.e., it satisfies%
\begin{equation}
N(\mathcal{W})=\sum\limits_{\mathbf{x}\in N}I(\mathbf{x}\in \mathcal{W}).
\label{NumofzerosPP}
\end{equation}

Note that we do not concern ourselves with the multiplicity of a zero, but
rather we want to find their number and their locations over a given bounded
window $\mathcal{W}\in \mathcal{B}(\mathcal{X}),$ and do this
simultaneously. An important consideration here, is that the zero locations
will be assumed to be independent of each other. We discuss relaxing this
assumption, i.e., replace independence via conditional independence in
Section \ref{DepzerosSection}. We briefly discuss the classic choice of
random measure for independent events next.

\subsection{Poisson Point Processes}

\label{PoisCoxSection}The most important point process model for independent
events is the Poisson process (e.g., \cite{DaleyVereJones2005}, \cite%
{DaleyVereJones2008}, \cite{Illian2008}, \cite{Gelfand2010}, \cite%
{Chiuetal2013}, \cite{Diggle2013} and \cite{baddeley2016spatial}). It is
often referred to as being \textquotedblleft completely at
random\textquotedblright\ or as a point process with \textquotedblleft no
interactions\textquotedblright ,\ since the number of events (and the events
themselves) over disjoint sets are independent of each other.

In what follows, let $\mu _{p}$ denote Lebesgue measure over $\mathbb{M}%
^{p}. $ Consider a window of observation $\mathcal{W}\subset \mathbb{M}^{p}$%
, and suppose that we observe $n$ points (events) $\varphi _{n}=\{\mathbf{x}%
_{i}\}_{i=1}^{n}$ from a point process $N,$ with $\mathbf{x}_{i}\in \mathcal{%
W}.$ In order to model this collection of points we consider the
inhomogeneous Poisson point process (IPPP), which assumes that the random
variables (counting variables) $N(B)$, $B\subseteq \mathcal{W},$ are
distributed as Poisson and that the counts are independent over any finite
collection of disjoint regions.

In particular, the expected number of events $\Lambda (B)=E\left[ N(B)\right]
$, known as the intensity measure, is the mean of a Poisson random variable,
i.e., $N(B)\thicksim Pois(\Lambda (B)).$ When $\Lambda $ is absolutely
continuous with respect to $\mu _{p}$, an appeal to the Radon-Nikodym
theorem yields
\begin{equation}
\Lambda (B)=E\left[ N(B)\right] =\tint\limits_{B}\lambda (\mathbf{x})\mu
_{p}(d\mathbf{x}),  \label{IntensityMeasure}
\end{equation}%
where $\lambda (\mathbf{x})$ is known as the intensity function and it
uniquely determines the distribution of the IPPP $N$. The special case where
$\lambda (\mathbf{x})=\lambda _{1}$ yields the homogeneous Poisson point
process (HPPP), with intensity $\lambda _{1},$ and mean measure $\Lambda
(B)=\lambda _{1}\mu _{p}(B),$ where $\mu _{p}(B)$ the volume of $B$. We
write $N\thicksim IPPP(\Lambda )$ or $N\thicksim IPPP(\lambda )$ to denote
this random measure.

The joint distribution of the independent events $\varphi _{n},$ and the
number of events $N(\mathcal{W})=n$ over the window $\mathcal{W},$ is given
by%
\begin{equation}
f_{N}(\varphi _{n},n)=\frac{e^{-\Lambda (\mathcal{W})}}{n!}%
\tprod\limits_{i=1}^{n}\lambda (\mathbf{x}_{i})=\frac{1}{n!}%
e^{-\tint\limits_{\mathcal{W}}\lambda (\mathbf{x})\mu _{p}(d\mathbf{x}%
)}\tprod\limits_{i=1}^{n}\lambda (\mathbf{x}_{i}),  \label{PoissonLik}
\end{equation}%
$n\in \mathbb{N}.$ Note that the intensity function uniquely determines the
IPPP, where $\lambda (\mathbf{x})\mu _{p}(d\mathbf{x})$ assumes the
interpretation of the probability of finding a point in an infinitesimal
ball of volume $\mu _{p}(d\mathbf{x})$ centered at $\mathbf{x}$.

Clearly, in order to describe the point process, one needs to model its
intensity $\lambda (\mathbf{x}),$ where the integral in Equation (\ref%
{IntensityMeasure}) is typically not available in closed form, requiring
numerical approximations, which in turn introduces a computational burden.
The standard approximation is via a Riemann sum; we grid the window $%
\mathcal{W}$ into $L$ fine, equal-sized cells, each with area $A_{l},$ and
label the centers of these cells as $\{\mathbf{u}_{l}:l=1,2,\dots ,L\}.$
Then we can approximate the integral using its Riemann sum%
\begin{equation}
\Lambda (\mathcal{W})=E\left[ N(\mathcal{W})\right] =\tint\limits_{\mathcal{W%
}}\lambda (\mathbf{x})\mu _{p}(d\mathbf{x})\simeq
\tsum\limits_{l=1}^{L}A_{l}\lambda (\mathbf{u}_{l}).
\label{EqApproxPPPIntegral}
\end{equation}%
Alternatively, we can use Monte Carlo integration; let $\mathbf{x}%
_{l}\thicksim Unif(\mu _{p}(\mathcal{W})),$ $l=1,2,\dots ,L,$ independent,
uniformly drawn points over $\mathcal{W}$, and write%
\begin{equation*}
\Lambda (\mathcal{W})=\tint\limits_{\mathcal{W}}\lambda (\mathbf{x})\mu
_{p}(d\mathbf{x})=\mu _{p}(\mathcal{W})\tint\limits_{\mathcal{W}}\lambda (%
\mathbf{x})\frac{1}{\mu _{p}(\mathcal{W})}\mu _{p}(d\mathbf{x})\simeq \frac{%
\mu _{p}(\mathcal{W})}{L}\tsum\limits_{l=1}^{L}\lambda (\mathbf{x}_{l}),
\end{equation*}%
where the Law of Large Numbers guarantees%
\begin{equation}
\frac{\mu _{p}(\mathcal{W})}{L}\tsum\limits_{l=1}^{L}\lambda (\mathbf{x}_{l})%
\overset{a.s.}{\rightarrow }\Lambda (\mathcal{W}),  \label{MonteCarloIntMeas}
\end{equation}%
as $L\rightarrow \infty .$

An exact method is obtained if we can write $\lambda (\mathbf{x})=\lambda
_{1}g(\mathbf{x}),$ where $g$ is a proper density over $\mathcal{W}$, and $%
\lambda _{1}>0$, some constant. Then we can rewrite Equation (\ref%
{IntensityMeasure}) as%
\begin{equation*}
\Lambda (\mathcal{W})=\tint\limits_{\mathcal{W}}\lambda _{1}g(\mathbf{x})\mu
_{p}(d\mathbf{x})=\lambda _{1}\tint\limits_{\mathcal{W}}g(\mathbf{x})\mu
_{p}(d\mathbf{x})=\lambda _{1},
\end{equation*}%
and therefore $\lambda _{1}$ is a constant describing the average number of
events over the region $\mathcal{W}$, and approximation of the integral is
no longer required. More details on this construction for the intensity
function can be found in \cite{ChakrabortyGelfand}, \cite{Micheas2014}, \cite%
{Micheas2019} and \cite{Micheas2025}. Next, we construct a specific IPPP to
help us find the zeros of a given function.

\subsection{Zeros via Poisson point processes}

In order to describe the zeros of $f\in \mathcal{F}_{\mathbb{M}}^{\mathbb{M}%
^{p}},$ we entertain an IPPP model for the random counting measure $N_{f}.$
Then, point patterns (realizations) from the IPPP $N_{f}$ will provide
realizations of the locations of the zeros. Recall that all we have to work
with is the function $f$, which will be assumed to satisfy the following.

\begin{remark}[Assumptions]
The function $f$ will be assumed to satisfy the following mild conditions.

\begin{enumerate}
\item $f$ is known in closed form or can be calculated/approximated well for
any $\mathbf{x}\in \mathbb{M}^{p}.$

\item $f$ has a finite number of zeros over a bounded set $\mathcal{W}$,
i.e., $N_{f}(\mathcal{W})<\infty .$

\item $f$ is continuous and uniformly bounded over a bounded set $\mathcal{W}%
.$

\item There does not exist a point $\mathbf{y}\in \mathbb{M}^{p},$ and
radius $r>0$, such that $f(\mathbf{x})=0,$ for all $\mathbf{x}$ in the open
ball $b(\mathbf{y},r).$
\end{enumerate}
\end{remark}

The latter condition is imposed in order to avoid degenerate or
uninteresting cases, e.g., when $f$ vanishes over a given interval, or it
oscillates between two values as $\mathbf{x}$ is rational or irrational, and
so forth.

By definition, the zeros satisfy $f(\mathbf{x})=0,$ $\mathbf{x}\in \Xi _{f},
$ so that in terms of an IPPP describing the random measure $N_{f}$ that
would yield a point pattern involving points from the zeros set $\Xi _{f},$
we must have an intensity function that would give less weight over all $%
\mathbf{x}\in \mathcal{W},$ when $f(\mathbf{x})\neq 0.$ Consequently, a
natural choice is to consider the IPPP $N_{f}$ with intensity function%
\begin{equation}
\lambda _{K,Q}(\mathbf{x})=\exp \left\{ -K|f(\mathbf{x})|^{Q}\right\} ,
\label{zerosIntfunc}
\end{equation}%
where $K,Q>0,$ and $|.|$ denotes absolute value of a real number when $%
\mathbb{M}=\mathbb{R}$ or modulus of a complex number when $\mathbb{M}=%
\mathbb{C}$. An appropriate norm must be employed when we depart from these
standard spaces.

Since an intensity measure that is absolutely continuous with respect to
Lebesgue measure assigns zero mass to every singleton. The intensity measure
$\Lambda $ is typically taken to be absolutely continuous with respect to
Lebesgue measure $\mu _{p},$ a nonatomic measure, and therefore it will not
be able to identify individual zeros.

Consequently, it is essential for the identification of individual candidate
locations, that the dominating measure $\mu $ of the intensity measure $%
\Lambda $ contains an atomic component. Specifically, define the integrating
measure%
\begin{equation}
\mu :=\mu _{p}+\sum\limits_{u\in \Xi _{f}}\delta _{u},  \label{mumeas}
\end{equation}%
with $\mu _{p}$ Lebesgue measure, a non-atomic measure on the continuous
part of the space, and $\delta _{u}$ Dirac measure, and note that $\mu
_{p}(\Xi _{f})=0.$

The following theorem encapsulates the theoretical properties of the PPPZ
construction.

\begin{theorem}
Consider the IPPP $N_{f}$ with intensity function (\ref{zerosIntfunc}) and
let $\Xi _{f}$ denote the set of zeros of $f$. The set $\Xi _{f}$ of zeros
of $f(\mathbf{x}),$ can be obtained as a realization of the IPPP $N_{f}$
with intensity function%
\begin{equation}
\lambda _{K,Q}(\mathbf{x})=e^{-K\left\vert f(\mathbf{x})\right\vert ^{Q}}I(%
\mathbf{x}\notin \Xi _{f})+I(\mathbf{x}\in \Xi _{f}),  \label{zerosIntfunc2}
\end{equation}%
as $K>0,$ fixed, increases. Moreover, with the integral understood as an
extended nonnegative integral, we have that the average number of zeros in
the window $\mathcal{W}\subseteq \mathcal{X}$ is given by%
\begin{equation}
\Lambda _{K,Q}(\mathcal{W})=\int\limits_{\mathcal{W}}\lambda _{K,Q}(\mathbf{x%
})\mu (d\mathbf{x})=\int\limits_{\mathcal{W}}e^{-K\left\vert f(\mathbf{x}%
)\right\vert ^{Q}}\mu _{p}(d\mathbf{x})+\#(\mathcal{W}\cap \Xi _{f}).
\label{intmeas}
\end{equation}%
In particular, if $\mathcal{W}\cap \Xi _{f}$ is finite, the second term is
its cardinality; if it is infinite, the integral is $+\infty $.\newline
Moreover, we have
\begin{equation}
\lim_{K\rightarrow +\infty }\lambda _{K,Q}(\mathbf{x})=\lambda _{\Xi _{f}}(%
\mathbf{x}):=I(\mathbf{x}\in \Xi _{f}),  \label{LimitIntens1}
\end{equation}%
and
\begin{equation}
\lim_{K\rightarrow +\infty }\Lambda _{K,Q}(\mathcal{W})=\lim_{K\rightarrow
+\infty }E\left[ N_{f}(\mathcal{W})\right] =\#(\mathcal{W}\cap \Xi _{f}).
\label{Numofzeros}
\end{equation}%
and the joint distribution of the zeros and their number over $\mathcal{W}$\
is given by%
\begin{equation}
f_{N_{f}}(\varphi _{n},n)\rightarrow f_{\Xi _{f}}(\varphi _{n},n)=\frac{1}{n!%
}e^{-\sum\limits_{\mathbf{\xi }\in \Xi _{f}}I(\mathbf{\xi }\in \mathcal{W}%
)}\tprod\limits_{i=1}^{n}\left[ \sum\limits_{\mathbf{\xi }\in \Xi
_{f}}\delta _{\mathbf{\xi }}(\mathbf{x}_{i})\right] ,  \label{LimDistr1}
\end{equation}%
as $K\mathbf{\rightarrow }+\infty .$ The IPPP with intensity function $%
\lambda _{\Xi _{f}}(\mathbf{x}),$ will be called the Poisson Point Process
of zeros (PPPZ).
\end{theorem}

\begin{proof}
\textbf{Connection to the zeros:} First we discuss how we can connect the
intensity function $\lambda _{K,Q}$ of the random measure $N_{f}$ with the
set of distinct zeros $\Xi _{f}$. Recall that the intensity function has the
following interpretation: if $\mu (d\mathbf{x})$ is the volume of an
infinitesimal sphere centered at $\mathbf{x}$, then $\lambda (\mathbf{x})\mu
(d\mathbf{x})$ is the probability that there is a point of the point process
in this sphere. Therefore, since we wish to describe the zeros of $f(\mathbf{%
x}),$ a natural choice is to consider the IPPP $N_{\xi }$ with intensity
function (\ref{zerosIntfunc}) which assigns probability 1 to a zero of $f(%
\mathbf{x}).$ In addition, we have%
\begin{equation*}
0<\lambda _{K,Q}(\mathbf{x})\leq 1,
\end{equation*}%
with the upper bound of $\lambda _{K,Q}(\mathbf{x})$\ attained at the zeros
of $f$ in $\mathcal{W}.$\newline
\textbf{Integrating measure: }In order to make $\lambda _{K,Q}(\mathbf{x})$
the intensity function of $N_{f}$, define the integrating measure $\mu $ as
in (\ref{mumeas}). We notice that for any measurable set $A$, we have%
\begin{equation*}
\mu (A)=\mu _{p}(A)+\sum\limits_{u\in \Xi _{f}}\delta _{u}(A)=\mu
_{p}(A)+\#(A\cap \Xi _{f}),
\end{equation*}%
where $\#(A\cap \Xi _{f})$ the cardinality of the set $A\cap \Xi _{f}$, and
further define%
\begin{equation*}
\mu (d\mathbf{x})=\mu _{p}(d\mathbf{x})+\sum\limits_{\mathbf{u}\in \Xi
_{f}}\delta _{\mathbf{u}}(d\mathbf{x}).
\end{equation*}%
\textbf{Explicit form of the intensity:} Then we can rewrite equation (\ref%
{zerosIntfunc}) as%
\begin{equation*}
\lambda _{K,Q}(\mathbf{x})=\exp \left\{ -K|f(\mathbf{x})|^{Q}\right\} I(%
\mathbf{x}\notin \Xi _{f})+e^{-K\times 0}I(\mathbf{x}\in \Xi _{f}),
\end{equation*}%
and since%
\begin{equation*}
I(\mathbf{x}\in \Xi _{f})=\sum\limits_{\mathbf{u}\in \Xi _{f}}I(\mathbf{u}=%
\mathbf{x}),
\end{equation*}%
we have%
\begin{equation}
\lambda _{K,Q}(\mathbf{x})=\exp \left\{ -K|f(\mathbf{x})|^{Q}\right\} I(%
\mathbf{x}\notin \Xi _{f})+\sum\limits_{\mathbf{u}\in \Xi _{f}}I(\mathbf{u}=%
\mathbf{x}).  \label{zerosIntfunc1}
\end{equation}%
\textbf{Intensity measure:} Write the intensity measure as an integral with
respect to the two parts of $\mu $ as%
\begin{equation*}
\Lambda _{K,Q}(\mathcal{W})=\int\limits_{\mathcal{W}}\lambda _{K,Q}(\mathbf{x%
})\mu _{p}(d\mathbf{x})+\sum_{u\in \Xi _{f}}\int\limits_{\mathcal{W}}\lambda
_{K,Q}(\mathbf{x})\delta _{\mathbf{u}}(d\mathbf{x}).
\end{equation*}

First consider the $\mu _{p}$-integral. Since $\Xi _{f}$ is countable and $%
\mu _{p}(\Xi _{f})=0$, the set $\Xi _{f}$ contributes nothing to this
integral. For $\mu _{p}$-almost every $\mathbf{x}\in \mathcal{W}$,
therefore, we have
\begin{equation*}
\lambda _{K,Q}(\mathbf{x})=e^{-K|f(\mathbf{x})|^{Q}},
\end{equation*}%
and hence
\begin{equation*}
\int\limits_{\mathcal{W}}\lambda _{K,Q}(\mathbf{x})\mu _{p}(d\mathbf{x}%
)=\int\limits_{\mathcal{W}}e^{-K|f(\mathbf{x})|^{Q}}\mu _{p}(d\mathbf{x}).
\end{equation*}

Next, for each $\mathbf{u}\in \Xi _{f}$, the Dirac measure satisfies
\begin{equation*}
\int\limits_{\mathcal{W}}\lambda _{K,Q}(\mathbf{x})\delta _{\mathbf{u}}(d%
\mathbf{x})=I(\mathbf{u}\in \mathcal{W})\lambda _{K,Q}(\mathbf{u}).
\end{equation*}%
At $\mathbf{x}=\mathbf{u}$, the first term in $\lambda _{K,Q}(\mathbf{x})$
vanishes because $\mathbf{u}\in \Xi _{f}$. In the sum of indicators, exactly
the term indexed by $\mathbf{u}$ equals $1$, while all other terms equal $0$%
. Thus
\begin{equation*}
\lambda _{K,Q}(\mathbf{u})=1,
\end{equation*}%
and it follows that the atomic contribution is
\begin{equation*}
\sum_{u\in \Xi _{f}}I(\mathbf{u}\in \mathcal{W})=\#(\mathcal{W}\cap \Xi
_{f}).
\end{equation*}

Adding the continuous and atomic contributions gives
\begin{equation*}
\Lambda _{K,Q}(\mathcal{W})=\int\limits_{\mathcal{W}}e^{-K|f(\mathbf{x}%
)|^{Q}}\mu _{p}(d\mathbf{x})+\#(\mathcal{W}\cap \Xi _{f}),
\end{equation*}%
as claimed.\newline
\textbf{Limiting intensity function: }If $\mathbf{x}\in \Xi _{f}$, then $f(%
\mathbf{x})=0$. The exponential term in
\begin{equation*}
\lambda _{K,Q}(\mathbf{x})=e^{-K|f(\mathbf{x})|^{Q}}I(\mathbf{x}\notin \Xi
_{f})+\sum_{\mathbf{u}\in \Xi _{f}}I(\mathbf{u}=\mathbf{x}),
\end{equation*}%
is zero because $\mathbf{x}\in \Xi _{f}$, while the sum of indicators is $1$%
. Therefore $\lambda _{K,Q}(\mathbf{x})=1$ for every $K>0$.

If $\mathbf{x}\notin \Xi _{f}$, then $f(\mathbf{x})\neq 0$, and the sum of
indicators is zero. Thus
\begin{equation*}
\lambda _{K,Q}(\mathbf{x})=e^{-K|f(\mathbf{x})|^{Q}}\longrightarrow 0,
\end{equation*}%
as $K\rightarrow +\infty .$ Together these cases give
\begin{equation*}
\lim_{K\rightarrow +\infty }\lambda _{K,Q}(\mathbf{x})=I(\mathbf{x}\notin
\Xi _{f}).
\end{equation*}%
\textbf{Limiting intensity measure: }Since $\Xi _{f}$ is countable, it has
Lebesgue measure zero. First note that the continuous contribution to $%
\Lambda _{K,Q}(\mathcal{W})$ is the integral
\begin{equation*}
\int\limits_{\mathcal{W}}e^{-K|f(\mathbf{x})|^{Q}}\mu _{p}(d\mathbf{x}).
\end{equation*}%
For Lebesgue-almost every $\mathbf{x}\in \mathcal{W}$, we have $\mathbf{x}%
\notin \Xi _{f}$, hence $f(\mathbf{x})\neq 0$, so the integrand converges to
$0$. It is bounded by $1$, and $\mathcal{W}\subseteq \mathcal{X}$ has finite
Lebesgue measure. Appealing to the dominated convergence theorem, we have
\begin{equation*}
\lim_{K\rightarrow +\infty }\int\limits_{\mathcal{W}}e^{-K|f(\mathbf{x}%
)|^{Q}}\mu _{p}(d\mathbf{x})=0.
\end{equation*}

At every $\mathbf{u}\in \Xi _{f}$, $\lambda _{K,Q}(\mathbf{u})=1$, so the
atomic contribution is, for every $K>0$, given by%
\begin{equation*}
\sum_{\mathbf{u}\in \Xi _{f}}\int\limits_{\mathcal{W}}\lambda _{K,Q}(\mathbf{%
x})\delta _{\mathbf{u}}(d\mathbf{x})=\sum_{\mathbf{u}\in \Xi _{f}}I(\mathbf{u%
}\in \mathcal{W)}=\#(\mathcal{W}\cap \Xi _{f}).
\end{equation*}%
Adding the continuous and atomic contributions and taking the limit proves
the result.\newline
\textbf{Joint distribution: }Finally, for the joint distribution of the
number of zeros $N_{f}(\mathcal{W})=\#(N_{f}\cap \mathcal{W})$ and their
locations $\varphi _{n}=\{\mathbf{x}_{1},...,\mathbf{x}_{n}\},$\ using
Equation (\ref{PoissonLik}) we have%
\begin{eqnarray*}
f_{N_{f}}(\varphi _{n},n) &=&\frac{1}{n!}e^{-\Lambda _{N_{f}}(\mathcal{W}%
)}\tprod\limits_{i=1}^{n}\lambda _{K,Q}(\mathbf{x}_{i})=\frac{1}{n!}%
e^{-\int\limits_{\mathcal{W}}e^{-K\left\vert f(\mathbf{x})\right\vert
^{Q}}\mu _{p}(d\mathbf{x})-\#(\mathcal{W}\cap \Xi _{f})} \\
&&\tprod\limits_{i=1}^{n}\left[ \exp \left\{ -K|f(\mathbf{x}%
_{i})|^{Q}\right\} I(\mathbf{x}_{i}\notin \Xi _{f})+\sum\limits_{\mathbf{\xi
}\in \Xi _{f}}\delta _{\mathbf{\xi }}(\mathbf{x}_{i})\right] ,
\end{eqnarray*}%
so that asymptotically, as $K\mathbf{\rightarrow }+\infty ,$ the joint
distribution of the number and locations of the IPPP $N_{f}$ has limit given
by Equation (\ref{LimDistr1}), which is the joint distribution $f_{\Xi
_{f}}(\varphi _{n},n)$ of the PPPZ.
\end{proof}

The PPPZ represents an IPPP with realizations being a subset of the fixed
set $\Xi _{f}.$ Since the PPPZ is constructed as the limit of the IPPP $%
N_{f} $, realizations of the latter can be used to approximate a subset of
the fixed set $\Xi _{f}.$ The other important aspect that is addressed in
the proposed construction, involves the distribution of the number of zeros
over the window of observation. The following result gives us some insight
on this important question, and it is a direct consequence of the underlying
point process framework we employed.

\begin{corollary}[Number of zeros]
Consider the IPPP $N_{f}$ with intensity function (\ref{zerosIntfunc}) and
let $\Xi _{f}$ denote the set of zeros of $f$. The number of zeros $N_{f}(%
\mathcal{W})=\#(N_{f}\cap \mathcal{W}),$ over a bounded window of
observation $\mathcal{W}$, is a Poisson random variable, i.e., $N_{f}(%
\mathcal{W})\thicksim Pois(\Lambda _{N_{f}}(\mathcal{W})),$ with mean and
variance satisfying%
\begin{equation*}
E\left[ N_{f}(\mathcal{W})\right] =Var\left[ N_{f}(\mathcal{W})\right]
=\Lambda _{N_{f}}(\mathcal{W})=\int\limits_{\mathcal{W}}\lambda _{K,Q}(%
\mathbf{x})\mu (d\mathbf{x})\rightarrow \#(\Xi _{f}\cap \mathcal{W}),
\end{equation*}%
as $K\rightarrow +\infty .$
\end{corollary}

Next we discuss special cases and generalizations of the new method for
finding zeros.

\subsection{Identifying Extrema and Generalizations}

First we note that the set of zeros $\Xi _{f}$ of a function $f$ coincides
with the set over which the anti-derivative of $f$, say $g(x)=\int f(x)d\mu
_{1}(x),$ with $f=g^{\prime },$ achieves all of its extrema (minima and
maxima) in $\mathcal{W}$. In\ other words, in order to find extrema of the
function $g$ we need to find the zeros of its derivative$\ f$. More
precisely, assume we wish to find the set where the extrema of $g:\mathbb{%
M\rightarrow M}$ are obtained, i.e.,
\begin{equation*}
E_{g}=\{\xi \in M:g(x)\leq g(\xi )\text{ or }g(\xi )\leq g(x),\text{ for all
}x\text{ in a neighborhood of }\xi \in \mathbb{M}\}.
\end{equation*}%
Then the global extrema over the window of observation $\mathcal{W}$\ are
obtained as%
\begin{equation*}
\xi _{\min }=\arg \underset{x\in E_{g}\cap \mathcal{W}}{\min }\{g(x)\},
\end{equation*}%
and%
\begin{equation*}
\xi _{\max }=\arg \underset{x\in E_{g}\cap \mathcal{W}}{\max }\{g(x)\}.
\end{equation*}

If the derivative $\frac{d}{dx}g(x)=g^{\prime }(x)=f(x)$ is available in
closed form, then the extrema of $g$ are obtained at the zeros $\Xi _{f}$ of
$f$. For multivariate functions $g:\mathbb{M}^{p}\mathbb{\rightarrow M},$ if
we have the gradient $\nabla g(\mathbf{x})=$ $\left[ \frac{\partial }{%
\partial x_{1}}g(\mathbf{x}),...,\frac{\partial }{\partial x_{p}}g(\mathbf{x}%
)\right] $ in closed form, we can use the PPPZ method for finding zeros of
the gradient function as we see next.

More generally, assume that we wish to find the set of zeros of a
multivariate, vector valued function $\mathbf{f}:\mathbb{M}^{p}\mathbb{%
\rightarrow M}^{q},$ $p,q\in \mathbb{N}^{+}=\{1,2,3,...\},$ with%
\begin{equation*}
\mathbf{f}(\mathbf{x})=[f_{1}(\mathbf{x}),...,f_{q}(\mathbf{x})],
\end{equation*}%
for all $\mathbf{x}\in \mathbb{M}^{p}.$ Now define the one dimensional
function of $\mathbf{x}$\ based on $\mathbf{f}$ given by%
\begin{equation*}
\left\Vert \mathbf{f}(\mathbf{x})\right\Vert =\sum\limits_{i=1}^{q}|f_{i}(%
\mathbf{x})|,
\end{equation*}%
so that $\mathbf{\xi }\in \Xi _{\mathbf{f}}=\left\{ \mathbf{x}\in \mathbb{M}%
^{p}:\mathbf{f}(\mathbf{x})=\mathbf{0}\right\} ,$ if and only if $\left\Vert
\mathbf{f}(\mathbf{\xi })\right\Vert =0$. Therefore, the PPPZ can be applied
to the modification of the intensity (\ref{zerosIntfunc}) given by%
\begin{equation}
\lambda _{K,Q}(\mathbf{x})=\exp \left\{ -K\left\vert
\sum\limits_{i=1}^{q}|f_{i}(\mathbf{x})|\right\vert ^{Q}\right\} ,
\label{MultivInt1}
\end{equation}%
and it will provide the zeros set $\Xi _{\mathbf{f}}$\ of $\mathbf{f}$.

When we wish to find all the points of extrema, i.e., local and global
minima and maxima, but it is hard to obtain the partial derivatives of $f$
in closed form, we consider finding the zeros of the approximation of the
derivative of $f$ given by%
\begin{equation}
\widehat{\frac{df(x)}{dx}}=\frac{f(x+\varepsilon )-f(x)}{\varepsilon },
\label{ApproxDeriv}
\end{equation}%
where $\varepsilon $ is very small. Similarly, in the multivariate case, we
use%
\begin{equation}
\widehat{\frac{\partial f(\mathbf{x})}{\partial x_{i}}}=\frac{f_{i}(\mathbf{x%
})-f(\mathbf{x})}{\varepsilon },  \label{ApproxDerivMultiv}
\end{equation}%
where%
\begin{equation}
f_{i}(\mathbf{x})=f(x_{1},...,x_{i-1},x_{i}+\varepsilon ,x_{i+1},...,x_{p}),
\end{equation}%
for each $i=1,2,...,p$. In order to find the extrema of $f$ we use $\widehat{%
\nabla f}(\mathbf{x})=$ $\left[ \widehat{\frac{\partial f(\mathbf{x})}{%
\partial x_{1}}},...,\widehat{\frac{\partial f(\mathbf{x})}{\partial x_{p}}}%
\right] $ to approximate $\nabla f(\mathbf{x})$ and then apply the PPPZ with
intensity function of Equation (\ref{MultivInt1})$,$ i.e., we take%
\begin{equation}
\lambda _{K,Q}(\mathbf{x})=\exp \left\{ -K\left\vert
\sum\limits_{i=1}^{p}\left\vert \widehat{\frac{\partial f(\mathbf{x})}{%
\partial x_{1}}}\right\vert \right\vert ^{Q}\right\} ,
\label{mutligradient1}
\end{equation}%
in order to find the solutions of $\nabla f(\mathbf{x})=\mathbf{0}.$

Finally, for any optimization problem or the PPPZs above, we can easily add
any constraints we wish directly to the intensity function. More precisely,
a constraint is described as a subset $\mathcal{C}$ of $\mathcal{X}$, e.g., $%
\mathcal{C}=\left\{ \mathbf{x}:\sum\limits_{i=1}^{p}x_{i}\geq 1\right\} ,$
and it simply alters the window of observation, i.e., we use the PPPZ with
intensity function $\lambda _{K,Q}(\mathbf{x})I(\mathbf{x}\in \mathcal{C})$
over the window $\mathcal{W}\subset \mathcal{X},$ where $I(.)$ denotes the
indicator function.

\section{Implementation considerations}

\label{Examples}The theoretical justification of the PPPZ method is
straightforward and intuitively appealing, however, in practice, since the
computers cannot handle infinity and even super computers overflow easily,
we will have to be very careful with the implementation of the proposed
method. Specifically, since we cannot perform a fine enough grid search to
check for a zero of $f$ (and we do not want to do that anyways), we will
have to produce a simulation method to approximate zeros that is a) as fast
as possible, b) accurate (up to our liking), and c) flexible enough to be
amenable to changes.

In particular, the PPPZ\ as constructed depends on the constants $K>0$ and $%
Q>0,$ which control how fast we will reject potential zeros, as well as, a
required tolerance variable $\tau $ that will act as the cutoff value for
what is an acceptable zero, i.e., when $|f(\mathbf{x})|<\tau,$ we will treat
$\mathbf{x}$ as a zero of $f$. For example, setting $\tau=10^{-500}$ will
never be easy to achieve in a reasonable amount of time and without
introducing overflows or memory allocation problems in the computer, or long
running times, even for easy, one dimensional functions. Instead, we will
provide a main simulation method to approximate the zero set $\Xi _{f}$ for
a specific starting $\tau_{start},$ and then if one wishes to get closer and
closer to the zero, we will propose an adaptive simulation method that can
take us to a reasonable target tolerance $\tau_{end}<\tau_{start},$ e.g., $%
\tau_{end}=10^{-10},$ before overflows occur in the computer. Moreover, as
we will see, the adaptive simulation method can take a lot of time to run
for $\tau_{end}<10^{-20}$, depending on the function under consideration and
the chosen values for parameters $K$ and $Q$.

It should be emphasized, that unlike existing methods for finding function
zeros or function optimization, such as simulated annealing or Newton
methods and their modifications, the proposed PPPZ can provide all zeros
using a single realization of the model (one iteration). Furthermore, the
method does not require the construction of a stochastic process that
converges to a zero or point of extrema, and neither requires forms for the
derivative of the function or iterations in order to obtain a zero. That
means the PPPZ does not suffer from any of the problems such methods are
notorious for; e.g., getting locked in local maxima or not achieving
convergence.

\subsection{Algorithms for finding zeros}

Simulating a HPPP $N$ with intensity $\lambda >0$ over a bounded set
(window) $\mathcal{W}$ is accomplished via conditioning, that is\newline
(i) simulate the number of points $N(\mathcal{W})\thicksim Pois(\lambda \mu
_{p}(\mathcal{W})),$ say $N(\mathcal{W})=n$, and then\newline
(ii) simulate $n$ points as an independent random sample from the uniform
distribution over $\mathcal{W}$.

In order to simulate the IPPP $N_{f}$ with intensity function of Equation (%
\ref{zerosIntfunc}), over a bounded set $\mathcal{X},$ we use a modification
of the standard thinning (rejection) sampler (see \cite{lewis1979simulation}%
). The following algorithm, and its adaptive version, were written in R,
incorporating faster C++ code via the RCPP and RcppArmadillo packages, and
are available to the scientific community, currently at this \href{https://sites.google.com/umsystem.edu/micheasa/research/function-zeros%
}{website}.

\begin{algorithm}[Poisson Point Process zeros]
\label{Algorithm1}Consider the IPPP $N_{f}\thicksim IPPP(\lambda _{K,Q}(.)),$
and note that for any given real $Q,K>0$, integer $N>1$ and any window $%
\mathcal{W}\subset \mathcal{X}\subset \mathbb{M}^{p},$ we have%
\begin{equation*}
\underset{\mathbf{x}\in \mathcal{W}}{\sup }\lambda _{K,Q}(\mathbf{x})=%
\underset{\mathbf{x}\in \mathcal{W}}{\sup }\exp \left\{ -K|f(\mathbf{x}%
)|^{Q}\right\} \leq 1<N.
\end{equation*}

\begin{enumerate}
\item Select a target tolerance $\tau >0;$ points $\mathbf{x}\in \mathcal{W}%
, $ with $|f(\mathbf{x})|<\tau ,$ are considered to be zeros of $f$. Note
that if this value is too small, it is possible that none of the zeros will
be identified from any realizations of $N_{f},$ no matter what the values of
$N$ or $K$.

\item {Simulate a $HPPP(N\mu _{p}(\mathcal{W}))$ over $\mathcal{W},$ say $%
\varphi _{n}=\{\mathbf{x}_{1},\ldots ,\mathbf{x}_{n}\}.$ Clearly, the larger
the }${N,}${\ the better the exploration of the window }$\mathcal{W}$, but
we do not want to perform a grid search, and therefore, $N$ will be set to $%
1000p$ or less, where $p$ is the dimension of the event space $\mathbb{M}%
^{p}.$

\item Independently retain each point $\mathbf{x}_{i}\in \mathcal{W},$ with
probability $\lambda _{K,Q}(\mathbf{x}_{i}),$ i.e., generate $U\thicksim
Unif(0,1),$ and retain the point if%
\begin{equation}
U<\exp \left\{ -K|f(\mathbf{x}_{i})|^{Q}\right\} .  \label{RetainProb}
\end{equation}

\item The realization of $N_{f}\thicksim IPPP(\lambda _{K,Q}(\mathbf{.})),$
consists of the retained events, and is treated as an approximation of the
set $\Xi _{f}.$
\end{enumerate}
\end{algorithm}

In order to achieve a specific given $\tau$ value for all zeros in Algorithm %
\ref{Algorithm1}, we consider the following modifications.

\begin{remark}[Adaptive PPPZ algorithm]
\label{AdaptiveAlgorithm}We discuss an adaptive version of Algorithm \ref%
{Algorithm1}, and the modification of each one of its steps, in order to
achieve a desired $\tau_{end}$. In the adaptive PPPZ (APPPZ) version, the
reader can choose a number of iterations, $Iter\geq 1.$ Without loss of
generality, assume $p=1$, and consider the zeros of $f$ over the window $%
\mathcal{W}_{1,1}=[x_{min},x_{max}]$.

\begin{enumerate}
\item The adaptive version of the algorithm, allows for a large starting
tolerance and a very small end tolerance. In particular, it is recommended
that the reader sets $\tau_{start}=10^{-1}$ and $\tau_{end}=10^{-10}.$ The
current tolerance $\tau_{cur}$ is set to $\tau_{start}.$

\item {The point pattern $\varphi _{n}$\ generated at this step can be
thought of as the starting value of the adaptive sampling algorithm.}

\item {The adaptive version of steps 3 and 4, looks first for zeros within
the current tolerance.}\newline
In order to achieve the specific tolerance $\tau_{cur},$ we retain a point if%
\begin{equation}
U<\exp \left\{ -K|f(\mathbf{x}_{i})|^{Q}\right\} I(|f(\mathbf{x}%
_{i})|<\tau_{cur}),
\end{equation}%
where $I(.)$ denotes the indicator function.\newline
For the first iteration only, {if no zeros are found in the realization of }$%
{N}_{{f}}${, the current tolerance is increased by }$10\tau_{cur},$ the
number of points is increased by $10N$, and we go back to step 1.\newline
Assume that we find {$\varphi _{Iter,n_{1}}=\{x_{1},\ldots ,x_{n_{Iter}}\}$}
potential zeros within $\tau_{cur}.$ We build smaller windows centered at
each of the potential zeros, $\mathcal{W}_{Iter,j}=[x_{j}-r,x_{j}+r],$ $%
j=1,2,...,n_{Iter},$ where $r$ is $1\%$ of the (parent) window $\mathcal{W}$%
, reduce the current tolerance by $10\%$, i.e., the next iteration tolerance
is $\tau_{cur}/10,$ and go back to step 1 for each window identified. Note
that increasing $r$ (up to $50\%$) will provide a better exploration about
the zero $x_{j}$ at the cost of computational speed.\newline
When $Iter>1,$ {if no zeros are found for the current window and given }$%
\tau_{cur}${, the window is slightly increased by setting }$1.1r,$ the
number of points is increased by $1.1N$, and we generate a new realization
of $N_{f} $ over the new window, until we get a zero within this tolerance
level.\newline
This procedure is guaranteed to refine the values where the zeros are
obtained for the current tolerance. The more the iterations, the closer to
the end tolerance we entertain the zeros will be, but the procedure becomes
very slow as the tolerance becomes smaller. Further note that since each
sub-window $\mathcal{W}_{Iter,j}$ can yield many realizations for a specific
zero within $\tau_{cur},$ e.g., say two zeros at $1.0001$ and $1.0002$, a
post processing effect is applied, where we keep only the zero with the
smallest $|f(\mathbf{x})|$ for all zeros identified within the window $%
\mathcal{W}_{Iter,j}.$

\item When the maximum number of iterations is reached or $%
\tau_{cur}<\tau_{end},$ the adaptive algorithm stops and the zeros within $%
\tau_{cur}$ are retrieved.\newline
Note that it is possible that the function $f$ does not have any zeros over
a given window. This case corresponds to none of the generated points being
valid zeros, i.e., satisfying $|f(\mathbf{x})|<\tau_{end},$ and it is
captured by the algorithm when a window yields no acceptable zeros. In this
case, the window "dies" out and is removed from further consideration.
\end{enumerate}
\end{remark}

The point process created by the adaptive algorithm of Remark \ref%
{AdaptiveAlgorithm}, is mathematically equivalent to a Neyman-Scott process
(for 2 iterations), where the parent process is the IPPP $N_{f}$ with
intensity function $\lambda _{K,Q}(.)$ of Equation (\ref{zerosIntfunc}), the
parents are removed, and in every window a child point process is generated
as an IPPP with intensity function $\lambda _{K,Q}(.)$, independently. The
Neyman-Scott process consists of the superposition of all offspring points.

For higher number of iterations, we generalize the Neyman-Scott process and
continue the following way; using the first level offspring as the parent
process, we generate the child processes at each window and obtain the
second level offspring point process as the superposition of all the
generated child processes. Proceed in a similar manner for any number of
iterations to obtain a finer and finer collection of offspring points, that
can get as close as we want to the zeros of $f$. When $\tau_{end}$ is
extremely small and the number of windows increases, we encountered three
major drawbacks; computational speed, computer overflows and memory
allocation problems.

\subsection{Choice of parameters $K$ and $Q$ in practice}

We provide some empirical and theoretical considerations when it comes to
the choice of parameters $K$ and $Q$ in practice.

In view of Equation (\ref{zerosIntfunc}), as the constant $Q>1$ becomes
larger, values of $|f(\mathbf{x})|$ that are less than $1$ become smaller
and smaller, so that larger $Q$ brings values of $|f(\mathbf{x})|<1$ closer
to zero even if there is no zero of $f$ at that point. As a consequence,
large $Q$ erroneously suggests that the intensity function $\lambda _{K,Q}$
is large at a specific point when it should not, and will lead to a
realization of the zero set $\Xi _{f}$ that may include non-zeros of the
function $f$. On the other hand, as $Q<1$ becomes smaller and smaller,
values of $|f(\mathbf{x})|<1$ become larger and larger, and the intensity
once again will erroneously suggest that there is no zero at a certain point
even if there is one there.

When $|f(\mathbf{x})|>1,$ the value $\mathbf{x}$ is certainly not a zero of $%
f$, and large values of $Q$ correctly increase $|f(\mathbf{x})|^{Q},$ so
that \ the intensity becomes smaller and this point will not appear in a
realization of $N_{f}$. Now as $Q\rightarrow 0^{+}$ from the right$,$ leads
to $|f(\mathbf{x})|^{Q}\rightarrow 1^{+},$ from the right, and even if we
have a zero at $\mathbf{x}$, it will not be retrieved in a realization of $%
N_{f}$, since the intensity will never go to zero.

The above observations suggest that appropriate values for the power should
be away from $0$ and not too large, e.g., $0.5\leq Q\leq 2.$ Following
extensive simulations, it is recommended that the readers use $Q=0.5$ for
real functions and $Q=2$ for complex functions. As the dimension increases
for multivariate real functions, it is best to choose $Q=1$.

Regarding the selection of $K>0$, and its effect in practice; although the
theory presented tells us that we must have $K\rightarrow \infty ,$ in
reality we cannot choose $K$ too large, since the checks required in
Equation (\ref{RetainProb}) will fail even at points that are zeros within
the given $\tau$\ value. As $K$ increases, the intensity will become smaller
and smaller even at points that are not the function zeros, which leads to
erroneously identifying zeros for the specific $\tau$\ value. Even if $%
\mathbf{x}_{i}$ is near a zero, with $|f(\mathbf{x}_{i})|<\tau,$ large $K$
values will erroneously reject this point from the realization of $N_{f}$.
On the other hand, as $K$ decreases, the intensity increases and the
algorithm will once again erroneously propose realizations that are
non-zeros of $f$. Therefore, in practice we need to use a relatively small $K
$, e.g., values $K=10,$ $20$ or $50$, seem to work best in simulated
examples. Setting $K=100,$ or larger, one encounters a situation where only
part of the set $\Xi _{f}$ appears in a realization of $N_{f}$, that is,
only part of the set of zeros are recovered. We provide some mathematical
justification and guidance on how parameter $K$ should be selected in the
following.

\begin{theorem}[Choice of $K$]
Consider the IPPP $N_{f}$ with intensity function (\ref{zerosIntfunc}) and
let $\Xi _{f}$ denote the set of zeros of $f$. Then a sufficient condition
for $\lambda _{K,Q}(\mathbf{x})\rightarrow \lambda _{\Xi _{f}}(\mathbf{x})$,
for all $\mathbf{x}\in \mathcal{W}$, is that for any finite window of
observation $\mathcal{W}$, and any fixed $Q>0,$ and small tolerance $0<\tau
<1,$ we have
\begin{equation}
K>K_{\tau ,Q}^{\ast }=-\frac{\log \left( \tau \right) }{M^{Q}},
\label{KBounds}
\end{equation}%
where $M=\underset{\mathbf{x}\notin \Xi _{f}\cap \mathcal{W}}{\max }|f(%
\mathbf{x})|>0.$ Specifically, when $0.5\leq Q\leq 2,$ the lower bound for $%
K $ becomes%
\begin{equation}
K>\left\{
\begin{tabular}{ll}
$-\frac{\log \left( \tau \right) }{\sqrt{M}},$ & $0<M<1$ \\
$-\log \left( \tau \right) ,$ & $M=1$ \\
$-\frac{\log \left( \tau \right) }{M^{2}},$ & $M>1$%
\end{tabular}%
\right. .  \label{Klowbound1}
\end{equation}
\end{theorem}

\begin{proof}
Using Equation (\ref{LimitIntens1}), $\forall \varepsilon >0,$ $\exists
K_{0}>0,$ such that $\forall K>K_{0},$ we have%
\begin{equation}
\left\vert \exp \left\{ -K|f(\mathbf{x})|^{Q}\right\} -\sum\limits_{\mathbf{%
\xi }\in \Xi _{f}}\delta _{\mathbf{\xi }}(\mathbf{x})\right\vert
<\varepsilon ,  \label{Limiepsdelta}
\end{equation}%
for all $\mathbf{x}\in \mathcal{W}$.\newline
When $\mathbf{x}\in \Xi _{f}\cap \mathcal{W},$ we have $f(\mathbf{x})=0,$
and $\sum\limits_{\mathbf{\xi }\in \Xi _{f}}\delta _{\mathbf{\xi }}(\mathbf{x%
})=1,$ so that Equation (\ref{Limiepsdelta}) is satisfied for any $K,Q>0$.%
\newline
Now choose an $\mathbf{x}\in \mathcal{W}$, such that $f(\mathbf{x})\neq 0,$
i.e., $\mathbf{x}\notin \Xi _{f}\cap \mathcal{W},$ which leads to $%
\sum\limits_{\mathbf{\xi }\in \Xi _{f}}\delta _{\mathbf{\xi }}(\mathbf{x}%
)=0. $ Choose $\varepsilon =\tau ,$ so that Equation (\ref{Limiepsdelta})
allows us to write%
\begin{equation*}
\exp \left\{ -K|f(\mathbf{x})|^{Q}\right\} <\tau ,
\end{equation*}%
and therefore%
\begin{equation*}
-K|f(\mathbf{x})|^{Q}<\log \left( \tau \right) .
\end{equation*}%
Solving for $K$ leads to the lower bound%
\begin{equation*}
K>-\frac{\log \left( \tau \right) }{|f(\mathbf{x})|^{Q}},
\end{equation*}%
which can be made independent of $\mathbf{x}$ by considering%
\begin{equation*}
0<m=\underset{\mathbf{x}\notin \Xi _{f}\cap \mathcal{W}}{\min }\left\vert f(%
\mathbf{x})\right\vert \leq \left\vert f(\mathbf{x})\right\vert \leq M=%
\underset{\mathbf{x}\notin \Xi _{f}\cap \mathcal{W}}{\max }|f(\mathbf{x})|,
\end{equation*}%
so that%
\begin{equation*}
\frac{1}{M^{Q}}\leq \frac{1}{|f(\mathbf{x})|^{Q}}\leq \frac{1}{m^{Q}},
\end{equation*}%
and therefore%
\begin{equation*}
K>-\frac{\log \left( \tau \right) }{M^{Q}}=K_{\tau ,Q}^{\ast },
\end{equation*}%
as entertained.\newline
Finally, since%
\begin{equation*}
\frac{dK_{\tau ,Q}^{\ast }}{dQ}=\frac{d}{dQ}\left( \frac{-\log \left( \tau
\right) }{M^{Q}}\right) =\frac{-\log \left( \tau \right) }{M^{Q}}\log \left(
\frac{1}{M}\right) =\frac{\log \left( \tau \right) \log (M)}{M^{Q}},
\end{equation*}%
for any $Q>0$ and $0<\tau <1,$\ we have%
\begin{equation*}
\frac{dK_{\tau ,Q}^{\ast }}{dQ}=\left\{
\begin{tabular}{ll}
$>0,$ & $0<M<1$ \\
$<0,$ & $M>1$%
\end{tabular}%
\right. .
\end{equation*}%
Specifically, for $0.5\leq Q\leq 2$ we have%
\begin{equation*}
-\frac{\log \left( \tau \right) }{\sqrt{M}}\leq K_{\tau ,Q}^{\ast }\leq -%
\frac{\log \left( \tau \right) }{M^{2}},
\end{equation*}%
for $0<M<1,$ and%
\begin{equation*}
-\frac{\log \left( \tau \right) }{M^{2}}\leq K_{\tau ,Q}^{\ast }\leq -\frac{%
\log \left( \tau \right) }{\sqrt{M}},
\end{equation*}%
for $M>1.$ When $M=1,$ we have trivially $K_{\tau ,Q}^{\ast }=-\log \left(
\tau \right) ,$ for any $Q>0.$ As a result, combining all these cases for $%
M>0$\ leads to (\ref{Klowbound1}) as required.
\end{proof}

In terms of sensitivity of the tolerance $\tau ,$ to the values of the lower
bound $K_{\tau ,Q}^{\ast }$ for $K,$ take $\tau =10^{-1},$ so that $-\log
(\tau )=2.302585,$ and as a consequence, for a function like $f(x)=\cos (x),$
with $M=1$, the lower bound for $K$ is $2.302585.$ Now as the tolerance
decreases, say $\tau =10^{-10},$ $10^{-15}$ or $10^{-20},$ then the lower
bound increases to $23.02585,$ $34.53878$ and $46.0517$, and this is not
surprising since $K_{\tau ,Q}^{\ast }=O(-\log \tau ).$ Therefore, the
empirical observations in this section are theoretically justified, and will
be further reinforced via the extensive simulation studies presented in the
next section.

\section{Applications}

In order to appreciate the effectiveness of the proposed methods, as well as
compare with existing methods of finding zeros and optimization, we present
some illustrative examples next.

\subsection{One dimensional real functions}

In the examples that follow we set $N=1000$, $K=10$, $Q=0.5$, $%
\tau_{start}=10^{-1}$ and $\tau_{end}=10^{-15}$. We begin with some simple
examples in order to appreciate the behavior of the PPPZ and Algorithm \ref%
{Algorithm1}, along with its modification of Remark \ref{AdaptiveAlgorithm}.

\begin{figure}[tbp]
\centering \includegraphics[width=0.495%
\textwidth]{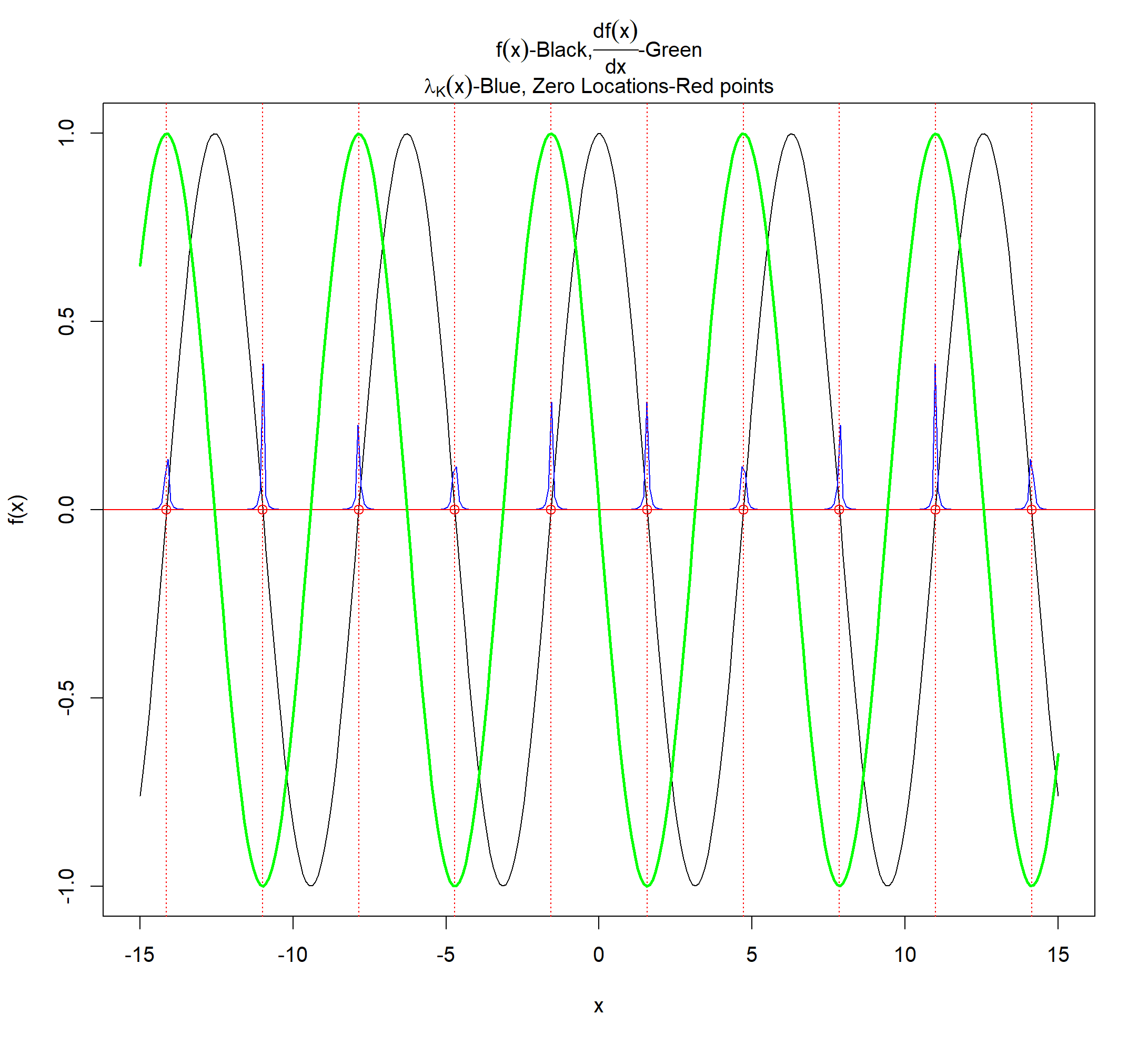} \includegraphics[width=0.495%
\textwidth]{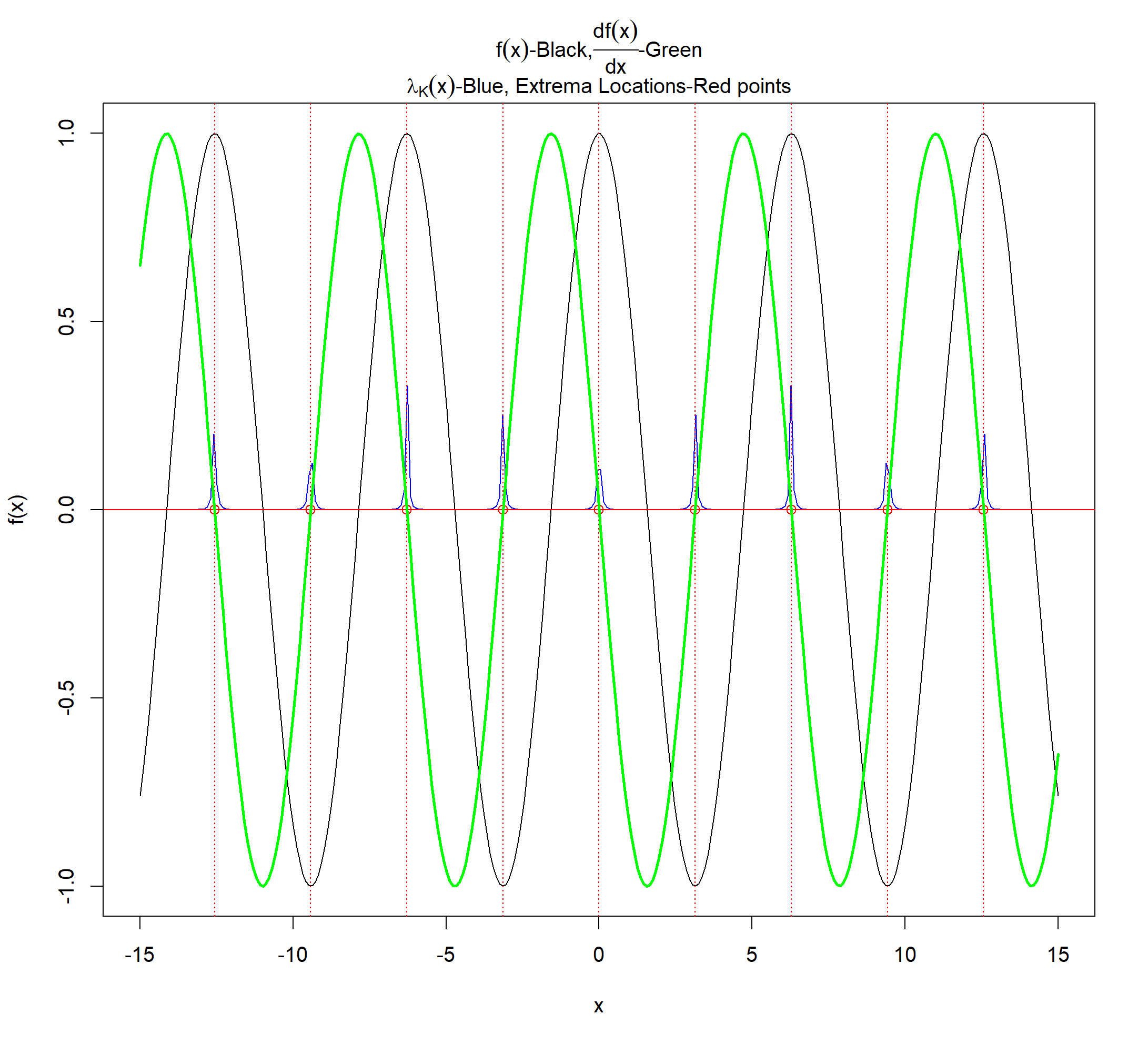}
\caption{The APPPZ algorithm over $\mathcal{W}=[-15,15]$ for the function $%
f(x)=cos(x)$ (black color). We run the algorithm for 1, 5, and 10
iterations, with the latter displayed (left plot). The derivative $%
f^{^{\prime}}(x)=-sin(x)$ is displayed in green and the intensity function
is displayed in blue color. The zeros are displayed as red points. Note that
the zeros of $cos(x)$ are the values where the extrema for the function $%
-sin(x)$ are attained. The zeros are presented in Table \protect\ref%
{CosineTable}. The right plot presents the results of the PPPZ on the zeros
of the derivative (i.e., the points of extrema of $cos(x)$).}
\label{CosMaxExamplesplot1}
\end{figure}

Consider the function $f(x)=\cos (x),$ with derivative $f^{\prime }(x)=-\sin
(x),$ over the window $\mathcal{W}=[-15,15]$. We run the APPPZ algorithm
(for 1, 5 and 10 iterations), and present the results in Table \ref%
{CosineTable} and Figure \ref{CosMaxExamplesplot1} (left plot, using results
for 10 iterations). The derivative is displayed in green and the intensity
function is displayed in blue color. The locations where the zeros are
obtained are displayed as red points. Note that the zeros of $\cos (x)$ are
the values where the extrema for the function $-\sin (x)$ are attained. The
zero values are presented in Table \ref{CosineTable}, where we can observe
the effect of the tolerance variable and the number of iterations, as well
as the time the PPPZ was run in each case.

We also run the APPPZ (10 iterations, 0.68s) on the derivative of $\cos (x)$
over $\mathcal{W}=[-15,15]$, in order to find the points of extrema. Figure %
\ref{CosMaxExamplesplot1} (right) presents the results where we found the 9
points of extrema (zeros of the derivative), with the global maximum being $1
$ attained at $x=-12.56637,$ $-6.283185,$ $2.200653e-11,$ $6.283185,$ $%
12.56637$, and the global minimum is $-1$ attained at $x=-9.424778,$ $%
-3.141593,$ $3.141593,$ and $9.424778$.

\begin{figure}[tbp]
\centering \includegraphics[width=0.495%
\textwidth]{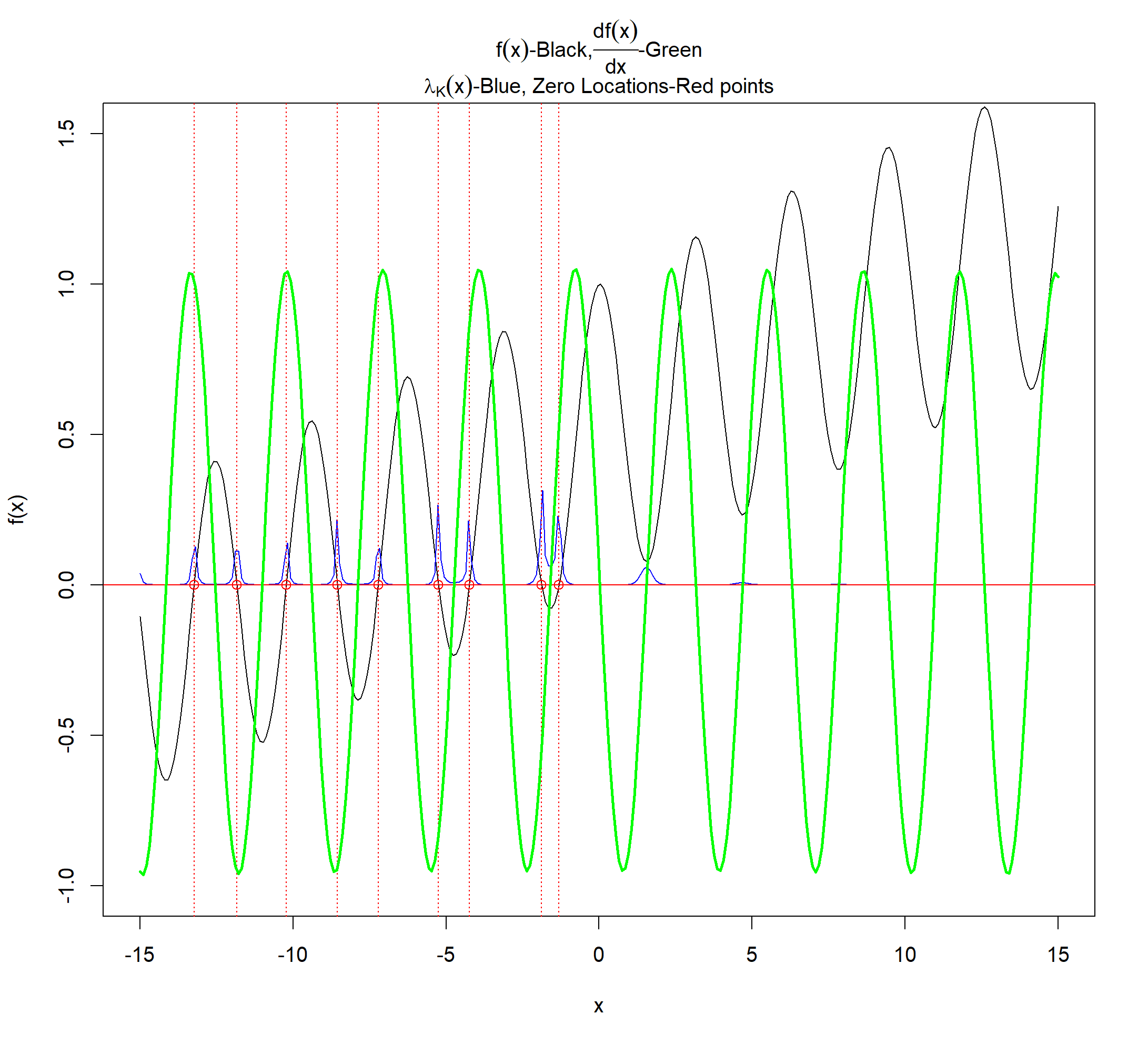} \includegraphics[width=0.495%
\textwidth]{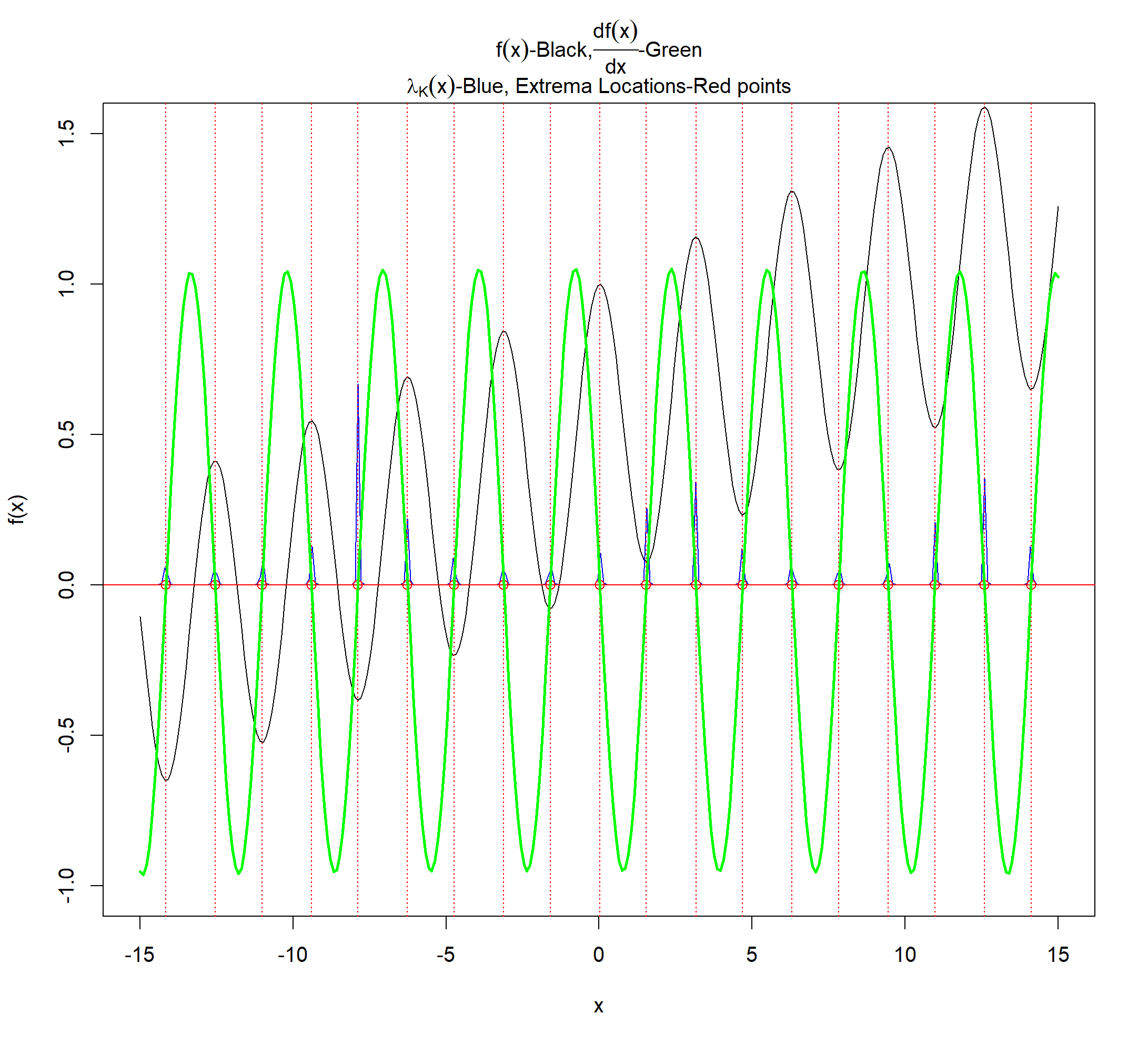}
\caption{The APPPZ algorithm over $\mathcal{W}=[-15,15]$ for the function $%
f(x)=\sin (x/20)+\cos ^{2}(x)$ (left plot in black color, algorithm was run
for 10 iterations). The right plot presents the results of the PPPZ on the
zeros of the derivative (i.e., the points of extrema of $\sin (x/20)+\cos
^{2}(x)$). The derivative is displayed in green and the intensity function
is displayed in blue color. The zeros are displayed as red points. }
\label{SinCosMaxExamplesplot1}
\end{figure}

Now consider the function $f(x)=\sin (x/20)+\cos ^{2}(x),$ over $\mathcal{W}%
=[-15,15],$ displayed in Figure \ref{SinCosMaxExamplesplot1} (left). This is
a more interesting function with a lot of different local maxima and local
minima. We run the APPPZ for 10 iterations (3.74s) and obtained 9 zeros
displayed in Table \ref{SinCosineExTable} (top). We also run the APPPZ (5
iterations, 0.32s) on the derivative of $\sin (x/20)+\cos ^{2}(x)$ in order
to find the points of extrema, with the results presented in Table \ref%
{SinCosineExTable} (bottom) and in Figure \ref{SinCosMaxExamplesplot1}
(right). We obtained 19 points of extrema with global maximum $1.588194$ at $%
x=12.58658$, and global minimum $-0.6498092$ at $x=-14.15617$.

\begin{figure}[tbp]
\centering \includegraphics[width=0.495\textwidth]{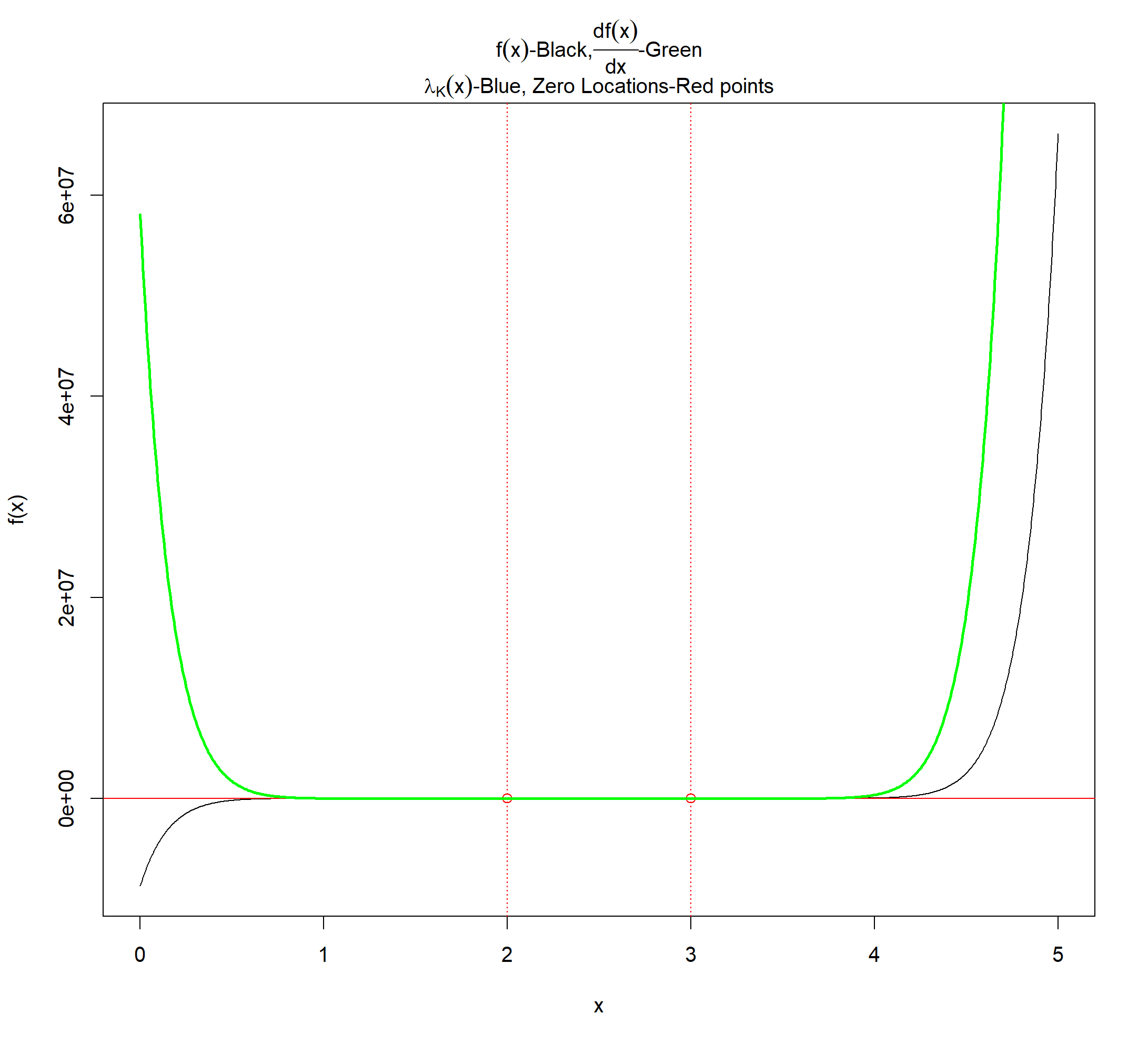}
\includegraphics[width=0.495\textwidth]{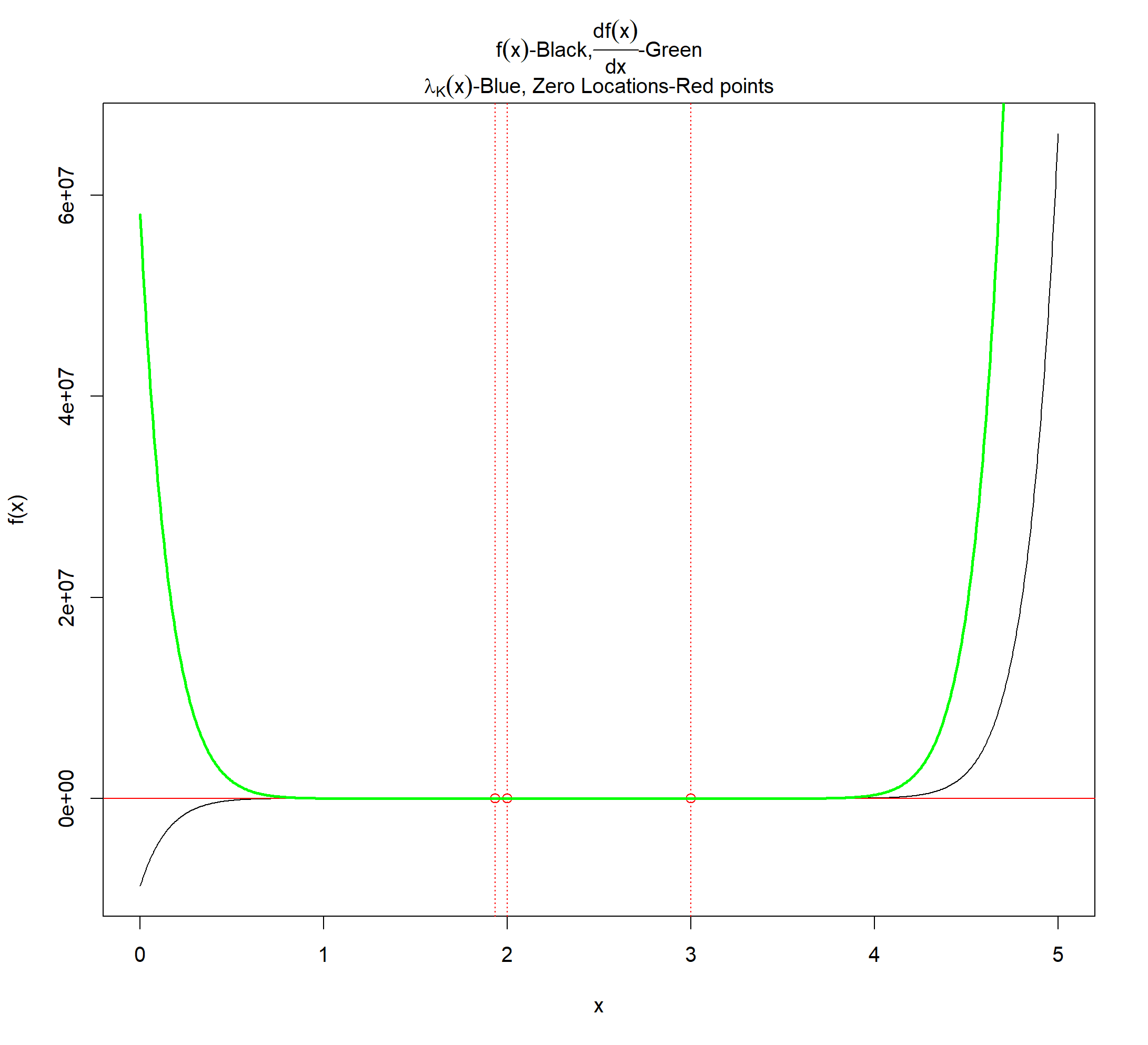}
\caption{The APPPZ algorithm over $\mathcal{W}=[0,5]$ for the function $%
f(x)=35(x-3)^{5}(x-2)^{10}$, in black color. We run the algorithm for 10
iterations in two cases with the left plot showing the true zeros at $x=2,3$%
, and the right plot illustrating what can happen when the function is very
close to zero for a range of its values; for the given tolerance $\protect%
\tau= 1e-10$, the adaptive algorithm ends up proposing three zeros. The
derivative is displayed in green and the intensity function is displayed in
blue color. The zeros are displayed as red points.}
\label{HardfuncExamplesplot1}
\end{figure}

Now consider the APPPZ algorithm over $\mathcal{W}=[0,5]$ for the function $%
f(x)=35(x-3)^{5}(x-2)^{10}$, displayed in Figure \ref{HardfuncExamplesplot1}
in black color. We run the algorithm for 10 iterations in two cases with the
left plot (55.14s) showing the true zeros at $x=2,3$, and the right plot
(61.01s) illustrating what can happen when the function is very close to
zero for a range of its values; for the given tolerance $\tau=10^{-10}$, the
adaptive algorithm ends up proposing three zeros. In particular, the two
zeros on the left plot correspond to $x=1.999900944$ with $%
f(x)=-3.184907493e-39$, and $x=2.999829027$ with $f(x)=-5.104621074e-18$,
while the three zeros in the right plot are obtained for $x=1.933949874$
with $f(x)=-7.615398755e-11$, $x=1.999994981$ with $f(x)=-3.553477504e-52$,
and $x=3.000340727$ with $f(x)=1.612789525e-16$. In such cases where the
function is very close to zero for a lot of values, the best approach is to
lower the tolerance and rerun the algorithm multiple times.

\begin{table}[tbp]
\centering {\tiny
\begin{tabularx}{\textwidth}{|XX|XX|XX|}
\hline
1 Iteration, 0.13s& $\tau= 0.1$ & 5 Iterations, 0.18s& $\tau= 1e-05$ & 10 Iterations, 0.49s& $\tau= 1e-10$\\
\hline
$x$ & $f(x)$&$x$ & $f(x)$&$x$ & $f(x)$\\
\hline
-14.137392144 &-2.252030286e-04&-14.137166819& 1.225845155e-07
&-14.137166941 &-1.093647799e-11\\
-10.995832121& 2.578329669e-04&-10.995573397& -8.904471795e-07
&-10.995574288&-1.534992806e-11\\
-7.854048223& -6.658926018e-05&-7.853984927&-3.292605847e-06
&-7.853981634&-8.302128345e-11\\
 -4.713078998& 6.900175898e-04&-4.712390497&1.516723439e-06
&-4.712388980&3.162873801e-11\\
-1.570671543& 1.247837813e-04&-1.570794105&2.222225828e-06
&-1.570796327&2.152972817e-11\\
1.571162785& -3.664586003e-04&1.570800172&-3.844757923e-06
&1.570796327&-4.959004857e-11\\
4.713414288& 1.025307562e-03&4.712387831& -1.148976112e-06
&4.712388980&4.585380358e-11\\
7.854884625& -9.029913153e-04&7.853985505 &-3.870612018e-06
&7.853981634&-1.200786608e-11\\
10.993944642&-1.629645175e-03&10.995574324&3.635807168e-08
&10.995574288&4.851187667e-11\\
14.134706296& 2.460642644e-03&14.137167606& -6.651373473e-07
&14.137166941&-1.299527556e-11\\
\hline
\end{tabularx}
}
\caption{The APPPZ algorithm for the function $f(x)=cos(x)$, after 1
iteration (left), 5 iterations (middle) and 10 iterations (right). Note that
we can get as close as we want to the zero, although even a single
realization (iteration) provides a good approximation to the zeros set $\Xi
_{f}$.}
\label{CosineTable}
\end{table}

\begin{table}[tbp]
\centering {\tiny
\begin{tabularx}{\textwidth}{|X|}
\hline
APPPZ zeros, 3.74s, $\tau=1e-10$, $x$ $(f(x))$\\
\hline
-13.236196796(7.433831328e-11),-11.839087124(-2.269140431e-11), -10.221080134(-9.227130171e-11),-8.553728223(-5.538963732e-11), \\
-7.217681823(1.911387715e-11), -5.246703175(-2.103101027e-11),
 -4.236080310(5.839356776e-12),-1.882374611(-2.198349836e-11),\\ -1.311891013(1.582591003e-11)\\
\hline
APPPZ extrema, 0.32s, $\tau= 1e-05$, $x$ $(f'(x))$ \\
\hline
-14.15616540 (1.755263e-06), -12.54612646 (3.221867e-06),
-11.01688533 (-4.938272e-06), -9.40247893 (-7.771561e-06),
-7.87707455 (2.540745e-06), -6.25938774 (-6.095124e-06),
-4.73670165 (-1.800227e-06),-3.11689054 (9.805490e-06),
-1.59573098 (-7.598505e-06), 0.02500947(1.765255e-06),1.54585904(-3.081008e-06),
3.16628790(4.198863e-06),4.68806496(3.904654e-06),
6.30696349(-4.112266e-06),7.83086437(-1.901257e-06),
9.44704644(2.697842e-06),10.97423862(1.572076e-06),
12.58658194(9.379164e-06),14.11814154(9.655610e-06)\\
\hline
\end{tabularx}
}
\caption{The APPPZ algorithm for the function $f(x)=\sin (x/20)+\cos ^{2}(x)$%
, to obtain the 9 function zeros (10 iterations, top), and the APPPZ for its
derivative to obtain the 19 points of extrema (5 iterations, bottom). }
\label{SinCosineExTable}
\end{table}

\subsection{One dimensional complex functions}

For the following examples we will assume that $p=1$ and $\mathbb{M}=\mathbb{%
C},$ the complex plane, and investigate classic complex-valued functions
such as polynomials with complex coefficients and Riemann's $\zeta $
function. In what follows, for any $s=\sigma +it\in \mathbb{C},$ we let $|s|=%
\sqrt{s\overline{s}}=\sqrt{\sigma ^{2}+t^{2}}$ denote complex modulus, where
$\overline{s}=\sigma -it\in \mathbb{C},$ the complex conjugate.

\subsubsection{Complex polynomials}

Define the complex polynomial $p(s)=(s-0.5+1i)^{2}(s-1.0-0.5i)^{3},$ for all
$s=\sigma +it\in \mathbb{C}$, with two roots at $s=0.5-1i$ and $s=1+0.5i.$
Consider the APPPZ with $N=1000$, $K=15$, $Q=2$, $\tau_{start}=10^{-1}$, $%
\tau_{end}=10^{-15}$, and intensity of the form%
\begin{equation}
\lambda _{K,Q}(s)=\exp \left\{ -K|p(s)|^{Q}\right\} =\exp \left\{ -15(p(s)%
\overline{p}(s))\right\} ,  \label{ComplexPoly1}
\end{equation}%
so that we require the zeros of the 2-dimensional real function $f(\sigma
,t)=f(s)=p(s)\overline{p}(s).$ The APPPZ algorithm was run for 10 iterations
(89.21s). We obtain two roots at $s=0.4999996331-1.0000000280i$, with
modulus square $f(s)=5.351451840e-13$, and $s=0.9999999858+0.5000061$ $853i$%
, with modulus square $f(s)=$ $5.915964743e-16$, so that the true roots are
recovered very well.

\begin{figure}[tbp]
\centering \includegraphics[width=0.495\textwidth]{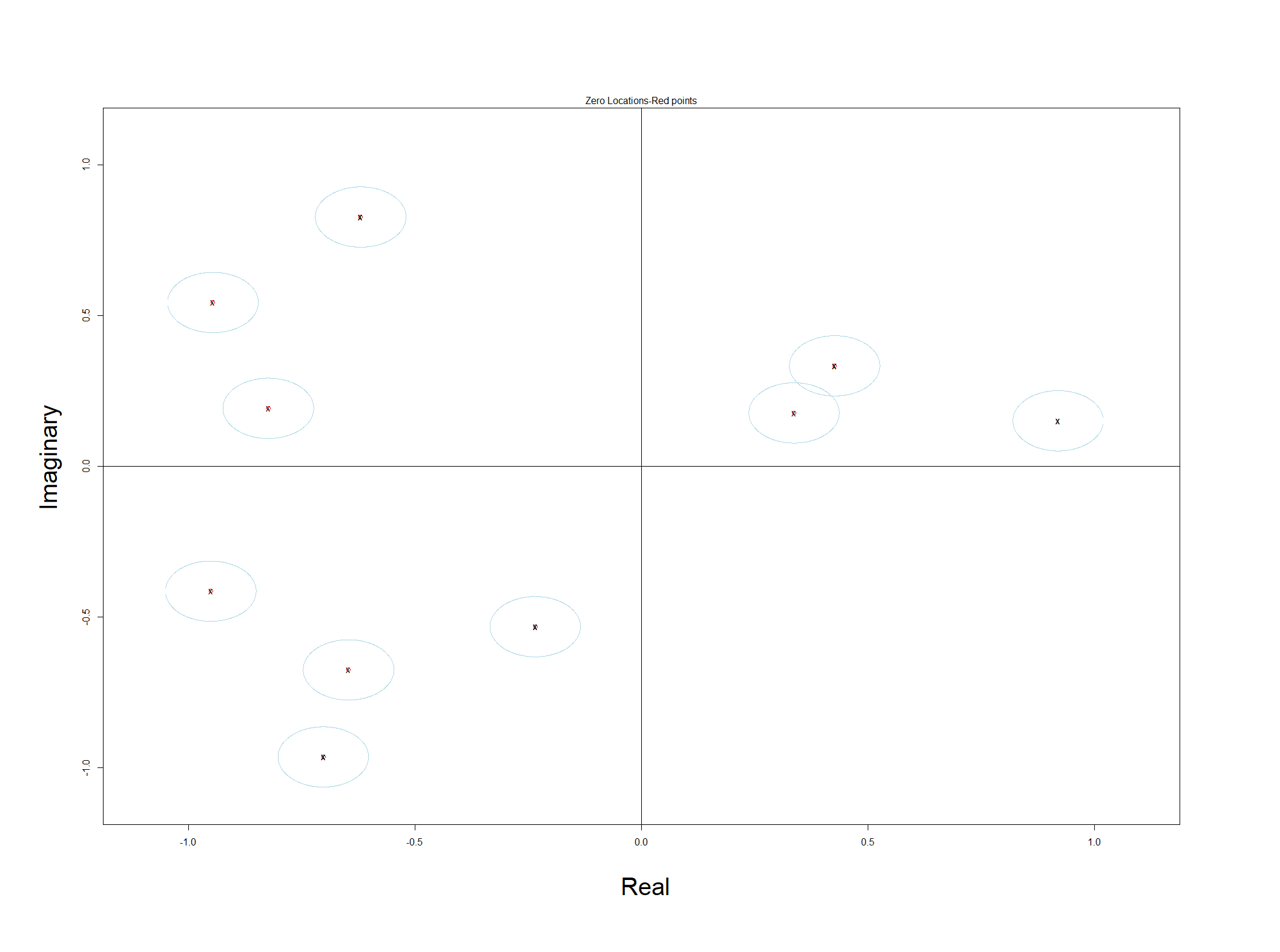} %
\includegraphics[width=0.495\textwidth]{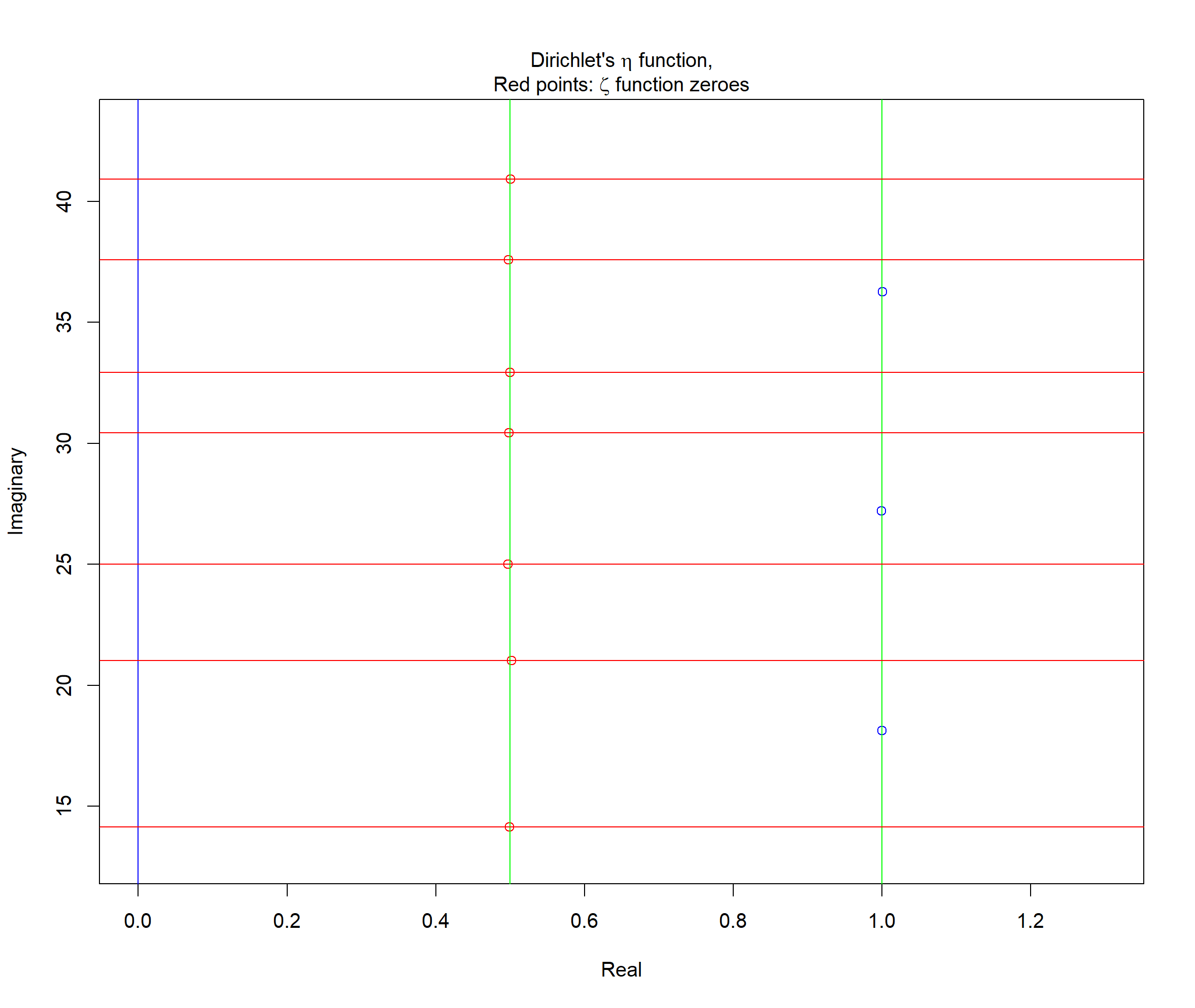}
\caption{(left) Roots of a complex polynomial $p_{10}(s)=\prod%
\limits_{k=1}^{10}(s-\protect\xi _{k}),$ with randomly generated real and
imaginary parts of the roots $\protect\xi _{k}$, independently, as uniform
random variables over $Unif(-1,1)$. The true roots denoted by x's (black)
and the recovered roots as o's (red). Note that all roots are recovered well
except for one (with real part near $0.9)$. \newline
(right) The APPPZ algorithm over $\mathcal{W}=\{\protect\sigma +it:0<\protect%
\sigma <1.3,13<t<43\}\subset \mathbb{C}$ for the $\protect\eta $ function.
The zeros of the $\protect\zeta$ function are displayed as red points and
the zeros of the $\protect\eta$ function on the strip $\protect\sigma=1$, as
green points. The red horizontal lines correspond to the well known zeros of
the strip $\protect\sigma=0.5$ with imaginary part $t=14.134725$, $21.022040$%
, $25.010858$, $30.424876$, $32.935062$, $37.586178$, and $40.918719$.}
\label{ZetaExamplesplot1}
\end{figure}

Now consider a complex polynomial $p_{n}(s)=\prod\limits_{k=1}^{n}(s-\xi
_{k}),$ where $s,\xi _{k}\in \mathbb{C},$ and $\xi _{k}$ are the roots, $%
s\in \mathcal{W}=\{\sigma +it:-1<\sigma <1,-1<t<1\}\subset \mathbb{C}$ . We
randomly generated the roots $\xi _{k}=\sigma _{k}+it_{k},$ $k=1,2,...,n$,
independently, as uniform random variables for their real and imaginary
parts, i.e., $\sigma _{k},t_{k}\thicksim Unif(-1,1)$. We take $n=10$ and run
the APPPZ (12584.62s) with $N=10000$, $K=10$, $Q=2$, $\tau_{start}=10^{-1}$,
$\tau_{end}=10^{-15}$, and intensity of the form (\ref{ComplexPoly1}). The
roots are displayed in Figure \ref{ZetaExamplesplot1} (left), with the true
roots denoted by x's (black) and the recovered roots as o's (red). Note that
all roots are recovered well except for one (with $\sigma $ near $0.9)$;
more precisely we have 9 roots at the following locations (with $p(s)%
\overline{p}(s)$ in parenthesis): $-0.9508137350-0.4145544947i$ $%
(3.134447927e-07)$, $-0.9460540885+0.5432386222i$ $(2.211998440e-06)$, $%
-0.8238205595+0.1920549400i$ $(3.662131736e-07)$, $%
-0.7024684132-0.9643830442i$ $(1.636778800e-06)$, $%
-0.6467435614-0.6756187027i$ $(1.572207974e-07)$, $%
-0.6199301022+0.8267882050 $ $(2.843105626e-06)$, $%
-0.2347544854-0.5322633364i$ $(9.451631995e-08)$, $%
0.3364580270+0.1769367759i $ $(4.945290$ $057e-07)$, and $%
0.4259629178+0.3331144697$ $(1.333234823e-06). $

\subsubsection{Riemann's $\protect\zeta $ function}

Consider Riemann's $\zeta $ function (see \cite{borwein2000efficient}, \cite%
{guillera2008double}) defined for any complex number $s=\sigma +it\in
\mathbb{C}$ by%
\begin{equation}
\zeta (s)=\sum\limits_{n=1}^{+\infty }\frac{1}{n^{s}}=\sum\limits_{n=1}^{+%
\infty }\frac{1}{n^{\sigma }}n^{-it}=\sum\limits_{n=1}^{+\infty }\frac{1}{%
n^{\sigma }}e^{-it\ln n},  \label{Zetafunc1}
\end{equation}%
where the sum converges if $\sigma >1.$ The $\zeta $ function is of
paramount importance in number theory, since it provides insight on the
distribution of prime numbers. The trivial zeros of $\zeta $ are obtained
when $s=-2,-4,-6,...,$ and any other zeros are called non-trivial zeros.

Riemann's hypothesis states that the real part of every non-trivial zero of
the $\zeta $ function is $0.5$, i.e., the remaining zeros are obtained on
the critical line $0.5+it$, for some $t\in \mathbb{R}$. Some of the zeros
include $1/2\pm 14.134725...i$, $1/2\pm 21.022040...i,$ $1/2\pm
25.010858...i,$ $1/2\pm 30.424876...i,$ $1/2\pm 32.935062...i,$ $1/2\pm
37.586178...i,$ and $1/2\pm 40.918719...i.$ Our purpose here is not to prove
or disprove the hypothesis, but rather illustrate the PPPZ for this
important function and verify some of the zeros that have been already
discovered in the literature.

Recall that Dirichlet's $\eta $ function (also known as the alternating $%
\zeta $ function, see \cite{sondow2003zeros}, \cite{guillera2008double}),
which converges for all $\sigma >0,$ is defined by%
\begin{eqnarray*}
\eta (s) &=&\sum\limits_{n=1}^{+\infty }\frac{(-1)^{n+1}}{n^{s}}%
=\sum\limits_{n=1}^{+\infty }\frac{(-1)^{n+1}}{n^{\sigma }}e^{-it\ln n} \\
&=&\left( \sum\limits_{n=1}^{+\infty }\frac{(-1)^{n+1}}{n^{\sigma }}\cos
(t\ln n)\right) -i\left( \sum\limits_{n=1}^{+\infty }\frac{(-1)^{n+1}}{%
n^{\sigma }}\sin (t\ln n)\right) ,
\end{eqnarray*}%
and therefore%
\begin{equation}
\eta (s)=\eta _{\func{Re}}(\sigma ,t)-i\eta _{\func{Im}}(\sigma ,t),
\label{Etafunc}
\end{equation}%
where%
\begin{equation}
\eta _{\func{Re}}(\sigma ,t)=\sum\limits_{n=1}^{+\infty }\frac{(-1)^{n+1}}{%
n^{\sigma }}\cos (t\ln n),  \label{EtaReal}
\end{equation}%
and%
\begin{equation}
\eta _{\func{Im}}(\sigma ,t)=\sum\limits_{n=1}^{+\infty }\frac{(-1)^{n+1}}{%
n^{\sigma }}\sin (t\ln n).  \label{EtaIm}
\end{equation}

Now note that the alternating form of the $\zeta $ function is defined via%
\begin{equation}
\zeta (s)=\frac{1}{1-2^{1-s}}\sum\limits_{n=1}^{+\infty }\frac{(-1)^{n+1}}{%
n^{s}}=\frac{1}{1-2^{1-s}}\eta (s),  \label{Zetafunc2}
\end{equation}%
and it is the analytic continuation of (\ref{Zetafunc1}) for all $\sigma >0,$
except for $\sigma =1.$ Considering the factor $1-2^{1-s},$ it is
straightforward to show that all zeros of $1-2^{1-s}$ are given by the
complex numbers $1+ik\pi /\log 2,$ for all integers $k\neq 0.$ Since%
\begin{equation}
\eta (s)=\left( 1-2^{1-s}\right) \zeta (s),  \label{EtaviaZeta}
\end{equation}%
for $\sigma >0,$ all zeros of $\eta $ include $1+i2k\pi /\log 2,$ $k\neq 0$,
and all the zeros of the $\zeta $ function. Therefore, assuming that
Riemann's hypothesis is correct, all zeros of $\eta $ for $0<\sigma <1,$ $%
t>0,$ must be located on the two parallel lines (strips) of the complex
plane with $\sigma =0.5$ and $\sigma =1.$

Consequently, we consider the PPPZ with intensity function (\ref%
{zerosIntfunc}) (with $K=15$, $Q=2$) of the form%
\begin{equation}
\lambda _{K,Q}(s)=\exp \left\{ -K|\eta (s)|^{Q}\right\} =\exp \left\{
-15(\eta (s)\overline{\eta }(s))\right\} ,  \label{EtazerosIntensity}
\end{equation}%
with $\overline{\eta }$ the conjugate of the function $\eta $, and using
Equation (\ref{Etafunc}), we have%
\begin{equation*}
\overline{\eta }(s)=\sum\limits_{n=1}^{+\infty }\frac{(-1)^{n+1}}{n^{\sigma }%
}e^{it\ln n}=\eta _{\func{Re}}(\sigma ,t)+i\eta _{\func{Im}}(\sigma ,t),
\end{equation*}%
for all $s\in \mathbb{C}$. Thus we can write the modulus square $|\eta
(s)|^{2}=\eta (s)\overline{\eta }(s)$ as%
\begin{equation*}
|\eta (s)|^{2}=\eta (s)\overline{\eta }(s)=\eta _{\func{Re}}^{2}(\sigma
,t)+\eta _{\func{Im}}^{2}(\sigma ,t),
\end{equation*}%
which a real function in two dimensions.

We run the APPPZ algorithm over the window $\mathcal{W}=\{\sigma
+it:0<\sigma <1.3,13<t<43\}\subset \mathbb{C}$ for the $\eta $ function with
the results displayed in Figure \ref{ZetaExamplesplot1} (right). The green
points correspond to the roots of the $\eta $ function on the strip $\sigma
=1,$ i.e., $0.9999948741+i18.13024288$ $(0.0010048257555),$ $0.9991999894+$ $%
i27.19393854$ $(0.0011775300376)$, and $1.0007632157+i36.25763092$ $%
(0.0016708950835)$. The non-trivial zeros of the $\zeta $ function are
displayed as red points at the locations $0.4994029097+i14.13722210$ $%
(0.0001929922065),$ $0.5022082300+$ $i21.02339177$ $(0.0013496548612),$ $%
0.4972257073+i25.01157078$ $(0.0016580104796),$ $0.4982$ $199803+$ $%
i30.42602791$ $(0.0010040109449)$, $0.4998933667+$ $i32.93208933$ $%
(0.0043031376534)$, $0.4975621602+$ $i37.58347713$ $(0.0046015734518)$, and $%
0.5006466505+i40.91937548$ $(0.0023$ $928247564)$.

A\ major drawback we encountered with the $\eta $ function is the
computational time it requires to get a good approximation of the function
values $\eta (s)$ for the given $s\in \mathbb{C}$. In particular, we used
the partial sums from Equations (\ref{EtaReal}) and (\ref{EtaIm}), given by%
\begin{equation*}
\eta _{\func{Re}}^{(L)}(\sigma ,t)=\sum\limits_{n=1}^{L}\frac{(-1)^{n+1}}{%
n^{\sigma }}\cos (t\ln n),
\end{equation*}%
and%
\begin{equation*}
\eta _{\func{Im}}^{(L)}(\sigma ,t)=\sum\limits_{n=1}^{L}\frac{(-1)^{n+1}}{%
n^{\sigma }}\sin (t\ln n).
\end{equation*}%
with $L=10000,$ so that the approximation of $\eta $ is given by%
\begin{equation*}
\widehat{\eta }(s)=\eta _{\func{Re}}^{(L)}(\sigma ,t)-i\eta _{\func{Im}%
}^{(L)}(\sigma ,t),
\end{equation*}%
which is not the most efficient way. More precisely, the results of Figure %
\ref{ZetaExamplesplot1} (right) were produced using one iteration of the
PPPZ for $N=10000000$, $K=15$, and required 30941.12s (8.59475hrs) to
complete. Note here that the choice of $L$ is directly affecting the
accuracy of the results of the PPPZ, i.e., for this $L$ we obtained the
zeros within tolerance $\tau=10^{-2}$.

\subsection{Multivariate real functions}

Consider the multivariate function $f(\mathbf{x})=\sum%
\limits_{i=1}^{p}(x_{i}-0.1)^{2},$ $\mathbf{x}=[x_{1},...,x_{p}]^{T}\in
\mathcal{W}$ $=\{\mathbf{x}:-1<x_{i}<1,i=1,2,...,p\}\subset \mathbb{R} ^{p},$
which has a single zero at the point $\mathbf{\xi }=[0.1,...,0.1]^{T}.$ We
investigate the behavior of the PPPZ for this function as the dimension $p$
increases. In Table \ref{MultivzerosEx} we present results of the APPPZ for $%
p=1,$ $2,$ $3,$ $4,$ and $5,$ (10 iterations, $Q=0.5$), and $p=10$ (5
iterations, $Q=1$). In all cases we used $K=10$, and $N=1000$. Note that all
zeros are recovered well, regardless of the specific dimension $p$.

\begin{table}[tbp]
\centering {\small
\begin{tabularx}{\textwidth}{|X|}
\hline
APPPZ zeros, $p=1$, 35.68s, $\tau=1e-10$, $x$ $(f(x))$\\
\hline
0.09999993692 (3.979408372e-15)\\
\hline
APPPZ zeros, $p=2$, 12.83s, $\tau=1e-10$, $\textbf{x}$ $(f(\textbf{x}))$\\
\hline
[0.1000009595, 0.1000012193] (2.407409327e-12)\\
\hline
APPPZ zeros, $p=3$, 21.23s, $\tau=1e-10$, $\textbf{x}$ $(f(\textbf{x}))$\\
\hline
[0.1000069522, 0.1000008998, 0.09999654371] (6.108910006e-11)\\
\hline
APPPZ zeros, $p=4$, 2.00s, $\tau=1e-10$, $\textbf{x}$ $(f(\textbf{x}))$\\
\hline
[0.1000019751, 0.1000017901, 0.1000052587,
 0.09999698607] (4.384352667e-11)\\
\hline
APPPZ zeros, $p=5$, 433.31s, $\tau=1e-10$, $\textbf{x}$ $(f(\textbf{x}))$\\
\hline
[0.09999492185, 0.09999744148, 0.09999579656, 0.09999903142, 0.1000030794] (6.042370764e-11)\\
\hline
APPPZ zeros, $p=10$, 2174.67s, $\tau=1e-4$, $\textbf{x}$ $(f(\textbf{x}))$\\
\hline
[0.1042333147, 0.09891379454, 0.09843709304, 0.09784912086, 0.09974686875, 0.1048708919, 0.09828548259, 0.1037441446, 0.1016098485, 0.1043649247] (8.856178743e-05)\\
\hline
\end{tabularx}
}
\caption{The APPPZ algorithm for the function $f(\mathbf{x}%
)=\sum\limits_{i=1}^{p}(x_{i}-0.1)^{2},$ $\mathbf{x}=[x_{1},...,x_{p}]\in
\mathbb{R}^{p},$ which has a single zero at the point $\mathbf{\protect\xi }%
=[0.1,...,0.1].$ Results for $p=1,$ $2,$ $3,$ $4,$ and $5,$ (10 iterations, $%
Q=0.5$), and $p=10$ (5 iterations, $Q=1$). In all cases we used $K=10$, and $%
N=1000$.}
\label{MultivzerosEx}
\end{table}

As an example of optimization, we illustrate maximum likelihood estimation
(MLE) of a multivariate normal model for different dimensions. More
precisely, consider the un-normalized, $p$-variate standard normal $\mathcal{%
N}_{p}(\mathbf{1},\mathbf{I}_{p})$, i.e., we will optimize the function%
\begin{equation*}
f(\mathbf{x})=\exp \left\{ -\frac{1}{2}\left\Vert \mathbf{x}-\mathbf{1}%
\right\Vert ^{2}\right\} ,
\end{equation*}%
for all $\mathbf{x}\in \mathcal{W}=\{\mathbf{x}:0<x_{i}<2,i=1,2,...,p\}%
\subset \mathbb{R}^{p},$ where $\mathbf{1}\in \mathbb{R}^{p},$ denotes the
unit vector $\mathbf{1}=[1,...,1]^{T}$. As rudimentary as this example is,
it provides additional insight on the performance of the APPPZ, and in
particular, when the partial derivatives are "not known" in closed form for
the multivariate, real-valued function under consideration. In Table \ref%
{MultivNormalMLEsEx} we present results of the APPPZ on the function $g(%
\mathbf{x})=\sum\limits_{i=1}^{p}\left\vert \widehat{\frac{\partial f(%
\mathbf{x})}{\partial x_{1}}}\right\vert $ (see Equation \ref{mutligradient1}%
)$,$ for $p=1,$ $2,$ $3,$ $4$ and $5,$ with $K=10$, $Q=1$, and $N=1000$. In
all cases, the single point of maximum is attained at the unit vector $%
\mathbf{\xi }=\mathbf{1},$ with the maximum value given by $f(\mathbf{\xi }%
)=1,$ as anticipated.

\begin{table}[tbp]
\centering {\small
\begin{tabularx}{\textwidth}{|X|}
\hline
APPPZ zeros, $p=1$, 20.61s, $\tau=1e-5$, $x$ $(g(x))$\\
\hline
0.9999999193 (2.997602166e-08)\\
\hline
APPPZ zeros, $p=2$, 55.93s, $\tau=1e-5$, $\textbf{x}$ $(g(\textbf{x}))$\\
\hline
[1.000000055, 1.000000257] (4.130029652e-07)\\
\hline
APPPZ zeros, $p=3$, 946.64s, $\tau=1e-5$, $\textbf{x}$ $(g(\textbf{x}))$\\
\hline
[1.000000484, 0.9999971188, 1.000002157] (5.573319584e-06)\\
\hline
APPPZ zeros, $p=4$, 10546.01s, $\tau=1e-04$, $\textbf{x}$ $(g(\textbf{x}))$\\
\hline
[0.9996077, 1.000057, 0.9995982, 0.9998747] (0.0009759304476)\\
\hline
APPPZ zeros, $p=5$, 7262.19s, $\tau=0.001$, $\textbf{x}$ $(g(\textbf{x}))$\\
\hline
[0.9981542, 0.9949539, 1.000481, 1.000696, 0.9982167] (0.00985266202)\\
\hline
\end{tabularx}
}
\caption{The APPPZ algorithm for the function $f(\mathbf{x})=\exp \left\{ -%
\frac{1}{2}\left\Vert \mathbf{x}-\mathbf{1}\right\Vert ^{2}\right\} ,$ $%
\mathbf{x}=[x_{1},...,x_{p}]\in \mathbb{R}^{p},$ which has a single maximum
at the unit vector $\mathbf{\protect\xi }=\mathbf{1}=[1,...,1]^{T},$ with
maximum value $f(\mathbf{\protect\xi })=1$. Results for $p=1,$ $2,$ $3,$ $4,$
and $5$. In all cases we used $K=10$, $Q=1$, and $N=1000$. }
\label{MultivNormalMLEsEx}
\end{table}

\begin{figure}[tbp]
\centering  \includegraphics[width=0.495\textwidth]{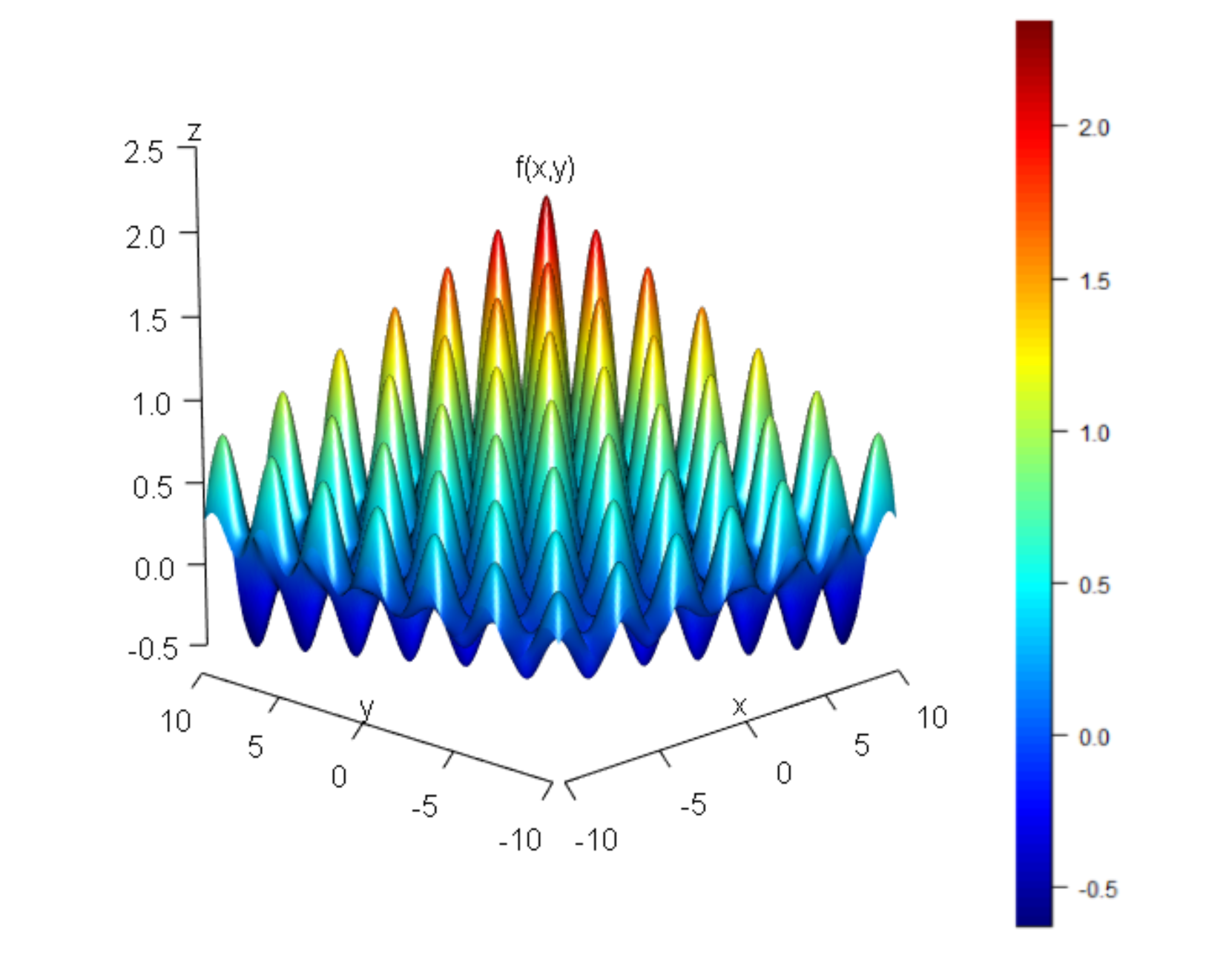} %
\includegraphics[width=0.495\textwidth]{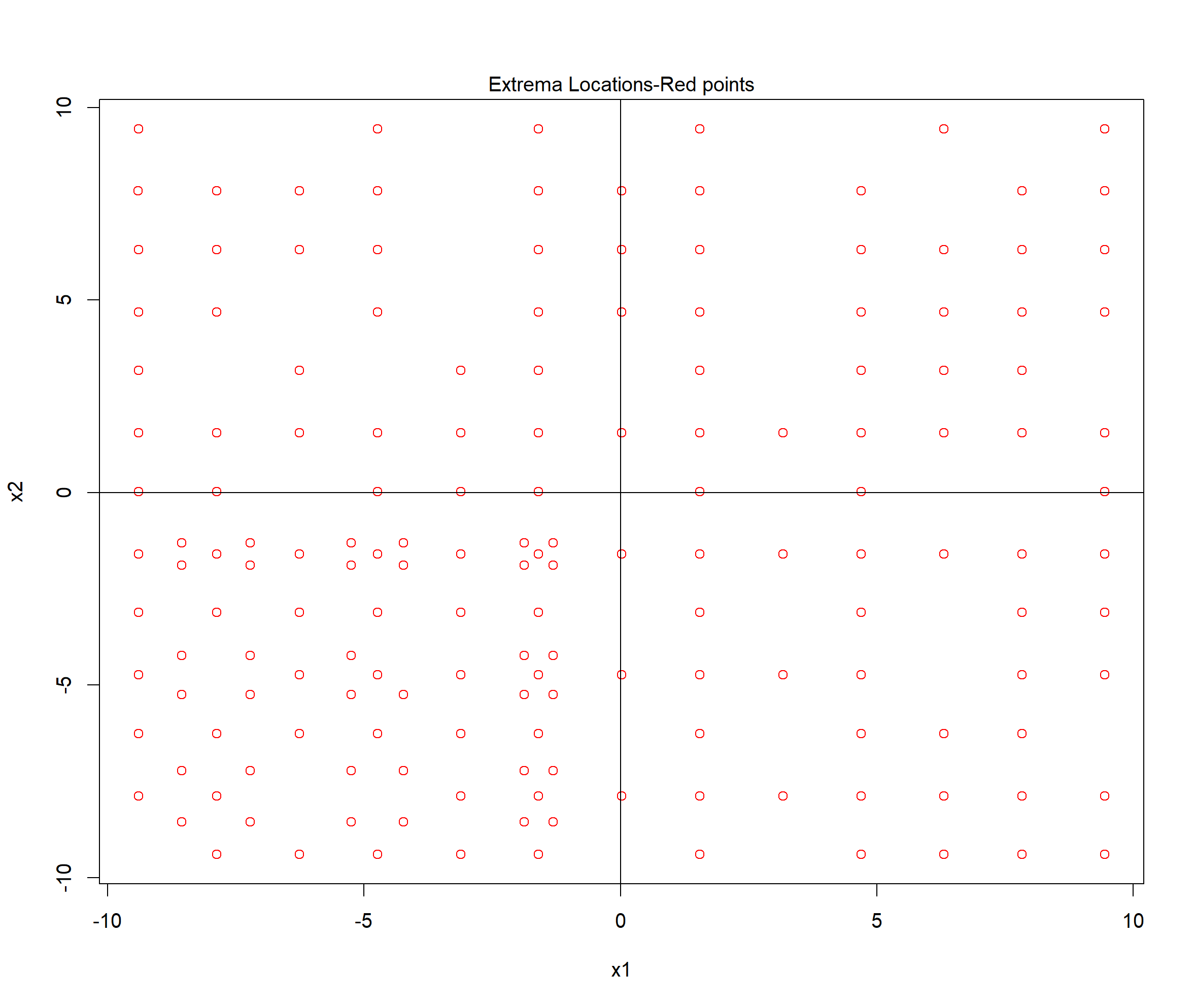}
\caption{(left) The function $f(x,y)=(\sin (x/20)+\cos ^{2}(x))(\sin
(y/20)+\cos ^{2}(y)),$ $(x,y)\in \mathcal{W}=[-10,10]^{2}\subset \mathbb{R}%
^{2}.$ (right) The APPPZ results from 5 iterations ($K=10$, $Q=0.5$, $N=1000,
$ $31.32483$hrs). All the points of extrema are displayed as red circles. }
\label{3dsincos1}
\end{figure}

Now consider the function $f(x,y)=(\sin (x/20)+\cos ^{2}(x))(\sin
(y/20)+\cos ^{2}(y)),$ $(x,y)\in \mathcal{W}=[-10,10]^{2}\subset \mathbb{R}%
^{2}.$ The surface is displayed in Figure \ref{3dsincos1} (left), where we
can see that the function attains its maximum at the point $\mathbf{\xi }%
=[9.438878,9.438878]^{T},$ but more importantly, there are many local minima
and maxima that present significant challenges to existing methods, such as
simulated annealing. We run the APPPZ for 5 iterations ($K=10$, $Q=0.5$, $%
N=1000,$ $31.32483$hrs) and all the points of extrema are displayed in
Figure \ref{3dsincos1} (right). Note that even in this nightmare scenario,
the APPPZ is able to recover almost all the points of extrema
simultaneously, with proposed global maximum at $[9.447049,9.447047]$ ($%
4.511946372e-06$) with $f(x,y)=2.115531$, and proposed global minimum at $%
[9.447045,-7.877075]$ ($3.871347687e-06$) with $f(x,y)=-0.5573836$.

\subsection{Comparisons with existing methods}

For comparison purposes, we used well known and established methods in order
to compare their performance against the PPPZ (PPPzeros R package) when it
comes to optimization. We will consider a relatively easy function, e.g., $%
f(x)=\sin (x/20)+\cos ^{2}(x),$ for $x\in \mathcal{W}=$ $[-15,15]$ (see
Figure \ref{SinCosMaxExamplesplot1}, left). We would expect existing methods
to do well in this trivial case, and we will assess if they outperform the
proposed PPPZ. Regardless, all existing methods in the literature, even if
they manage to provide the correct point where the global maximum is
attained, they do not provide the whole set of extrema as in the PPPZ
approach, which is one of the main advantages of the proposed method herein.
In all cases, we did not use the derivative (which is known in closed form),
but instead utilized the basic numerical approximation of the derivative
included in these R functions.

We utilize well known R packages and functions to perform the comparisons,
including the optimx and psoptim R packages, and the optim function from the
stats R package. In particular, we will attempt to replicate the
optimization via the APPPZ presented in Table \ref{SinCosineExTable}, using
several standard methods in the literature, such as box-constrained
optimization using PORT routines (nlminb, see \cite{gay1990usage}),
conjugate gradient algorithm (Rcgmin, see \cite{dai2001efficient}), limited
memory quasi-Newton algorithm (L-BFGS-B , see \cite{byrd1995limited}),
particle swarm optimization (PSO, see \cite{banks2007review}), and simulated
annealing (see \cite{belisle1992convergence}).

In Table \ref{ComparisonOptimTable} we present the comparison results. Note
that certain methods allow and are directly affected by the starting value
used in the algorithm. We randomly generated (uniformly) three starting
values from the window $[-15,15]$, given by $x_{0}=-2.642015$, $-0.1388709$,
and $13.73675$. For these methods (L-BFGS-B, nlminb, Rcgmin, and simulated
annealing), only when we start the algorithm near the global maximum $%
12.58658$ do these methods provide the correct answer. In contrast, the
APPPZ and PSO are able to reach the correct global maximum. The results
using the simulated annealing method where not surprising, since the
algorithm kept getting locked into local maxima and did not acquire the
global maximum unless it was started near the correct point. To be fair,
this function presents a very hard situation for the simulated annealing
approach when it comes to function maximization.

We further compared the APPPZ and PSO for the function $f(x,y)=(\sin
(x/20)+\cos ^{2}(x))(\sin (y/20)+\cos ^{2}(y)),$ $(x,y)\in \mathcal{W}%
=[-10,10]^{2}\subset \mathbb{R}^{2}$ (see Figure \ref{3dsincos1}, left).
Recall that the APPPZ correctly identifies the global maximum at $%
[9.447049,9.447047]$ with $f(x,y)=2.115531$. Unfortunately, the PSO did not
perform well when we increased the problem to two dimensions, even for this
simple bivariate function. We ran the algorithm for an ever increasing
number of particles, in an attempt to help the PSO method identify the
global maximum. The number of particles $n$, point of global maximum $%
\mathbf{x}_{0}$ and maximum value $f(\mathbf{x}_{0})$ are shown in Table \ref%
{ComparisonPSOOptimTable}. Note that not even once does the PSO identify the
correct point of global maximum.

\begin{table}[tbp]
\centering {\small
\begin{tabularx}{\textwidth}{|>{\hsize=0.25\hsize}X | >{\hsize=0.75\hsize}X|}
\hline
Method & Function $f(x)=\sin (x/20)+\cos ^{2}(x)$, $x\in[-15,15]$\\
\hline
 APPPZ&$x_{max}=12.58658$, $f(x_{max})=1.588194$\\
\hline
 L-BFGS-B\newline Quasi-Newton&
 $x_0=13.73675$, $x_{max}=12.58659$, $f(x_{max})=1.588194$\newline
 $x_0=-0.1388709$, $x_{max}=0.02501009$, $f(x_{max})=1.000625$\newline
 $x_0=-2.642015$, $x_{max}=-3.116886$, $f(x_{max})=0.8441755$\\
\hline
 Particle swarm&$x_{max}=12.58669$, $f(x_{max})=1.588194$\\
\hline
 nlminb& $x_0=13.73675$, $x_{max}=12.58659$, $f(x_{max})=1.588194$\newline
 $x_0=-0.1388709$, $x_{max}=0.02501041$, $f(x_{max})=1.000625$\newline
 $x_0=-2.642015$, $x_{max}=-3.116885$, $f(x_{max})=0.8441755$ \\
\hline
 Rcgmin& $x_0=13.73675$, $x_{max}=12.58659$, $f(x_{max})=1.588194$\newline
 $x_0=-0.1388709$, $x_{max}=0.02501041$, $f(x_{max})=1.000625$\newline
 $x_0=-2.642015$, $x_{max}=-3.116886$, $f(x_{max})=0.8441755$\\
\hline
 Simulated annealing & $x_0=13.73675$, $x_{max}=12.58659$, $f(x_{max})=1.588194$\newline
 $x_0=-0.1388709$, $x_{max}=0.02501009$, $f(x_{max})=1.000625$\newline
 $x_0=-2.642015$, $x_{max}=-3.116886$, $f(x_{max})=0.8441755$ \\
\hline
\end{tabularx}
}
\caption{Approximation to the global maximum for the function $f(x)=\sin
(x/20)+\cos ^{2}(x)$, within the window $[-15,15]$, based on several
established methods and the APPPZ (single realization with tolerance $%
\protect\tau=10^{-5}$, see the results in Table \protect\ref%
{SinCosineExTable}). The starting value $x_0=-2.642015$, $-0.1388709$, or $%
13.73675$, of an algorithm, when applicable, was randomly selected
(uniformly) from the window. See the text for more details.}
\label{ComparisonOptimTable}
\end{table}

\begin{table}[tbp]
\centering {\small
\begin{tabularx}{\textwidth}{|X|X|X|}
\hline
Number of particles $n$& Global maximum $\mathbf{x}_{0}$&
Value $f(\mathbf{x}_{0})$\\
\hline
 50&[-0.01498238, -3.132079]& 0.8431232\\
\hline
 500&[6.357042, 0.06729871]& 1.305569\\
\hline
5000&[9.42258, -9.982755]& 0.3503746\\
\hline
50000&[9.460585, -3.068967]& 1.22436\\
\hline
 500000&[-0.3548142, -3.108986]& 0.7272561\\
\hline
 5000000&[-0.02893543, -6.296969]& 0.6885615\\
\hline
 50000000&[3.056179, 5.945503]& 1.354638\\
\hline
\end{tabularx}
}
\caption{Approximation to the global maximum for the function $%
f(x,y)=(\sin(x/20)+\cos ^{2}(x))(\sin (y/20)+\cos ^{2}(y)),$ $(x,y)\in
\mathcal{W}=[-10,10]^{2}\subset \mathbb{R}^{2}$, using the PSO method with
several particles.}
\label{ComparisonPSOOptimTable}
\end{table}

\subsection{Dependent zeros}

\label{DepzerosSection}Throughout the exposition in this introductory paper
to the PPPZ we have made a major assumption about the zeros, that of
independence. We briefly discuss how we can relax this assumption, since
there are situations where this assumption may be unreasonable. For a
classic example of dependent zeros, take $\mathbb{M}=\mathbb{C},$ and define
the random power series of the form
\begin{equation*}
f_{\mathbf{U}}(z)=\sum\limits_{n=0}^{+\infty }a_{n}z^{n},
\end{equation*}%
where $\left\{ a_{n}\right\} _{n=0}^{\infty }$ are independent, standard
complex Gaussian random variables (with density $f(z)=e^{-z\overline{z}}/\pi
,$ $z\in \mathbb{C}$), and the radius of convergence of the series is a.s.
1. The set of zeros forms a point process $Z_{\mathbf{U}}$ in the unit disk $%
\mathbf{U}$, known as a determinantal process in $\mathbf{U}$ (see \cite%
{peres2005zeros}, \cite{shirai2025density} and the references therein for
the details).

In order to introduce dependent zeros in the proposed point process
framework we can consider the well known Cox random measures as follows.
Clearly, the choice of $K$ greatly influences the effectiveness of the APPPZ
algorithm. One approach that is currently being explored in order to refine
the proposed theory and algorithms (these results will be presented
elsewhere), involves treating $K$ as a random variable, e.g., take $%
K\thicksim Gamma(a,\beta ),$ $a,\beta >0.$ In this case, the intensity of
the IPPP of Equation (\ref{zerosIntfunc}) becomes a random intensity (a
random field), and the resulting random measure is known as a Cox point
process (see \cite{Micheas2025}, and the references therein).

A Cox random measure essentially replaces the independence assumption of the
Poisson point process with conditional independence, i.e., the numbers of
points over disjoint regions are not necessarily independent, and therefore,
it is appropriate to model dependence in the set $\Xi _{f}$. In Figure \ref%
{CoxExamplesplot1} we consider the zeros of $f(x)=\cos (x)$, over the window
$\mathcal{W}=[-10,10]$, and present 1000 realizations (left plot) of the
random intensity function $\lambda _{K,0.5}(\mathbf{x}),$ with source of
randomness the parameter $K\thicksim Gamma(5,2).$ The plot on the right
presents 95\% envelopes of the random intensity, as well as its mean. As
expected, the intensity is exemplified about the zeros of the function, and
the APPPZ algorithm provides a realization of $\Xi _{f}$ by assuming
conditional independence instead of independence of the zeros.
\begin{figure}[tbp]
\centering \includegraphics[width=0.495\textwidth]{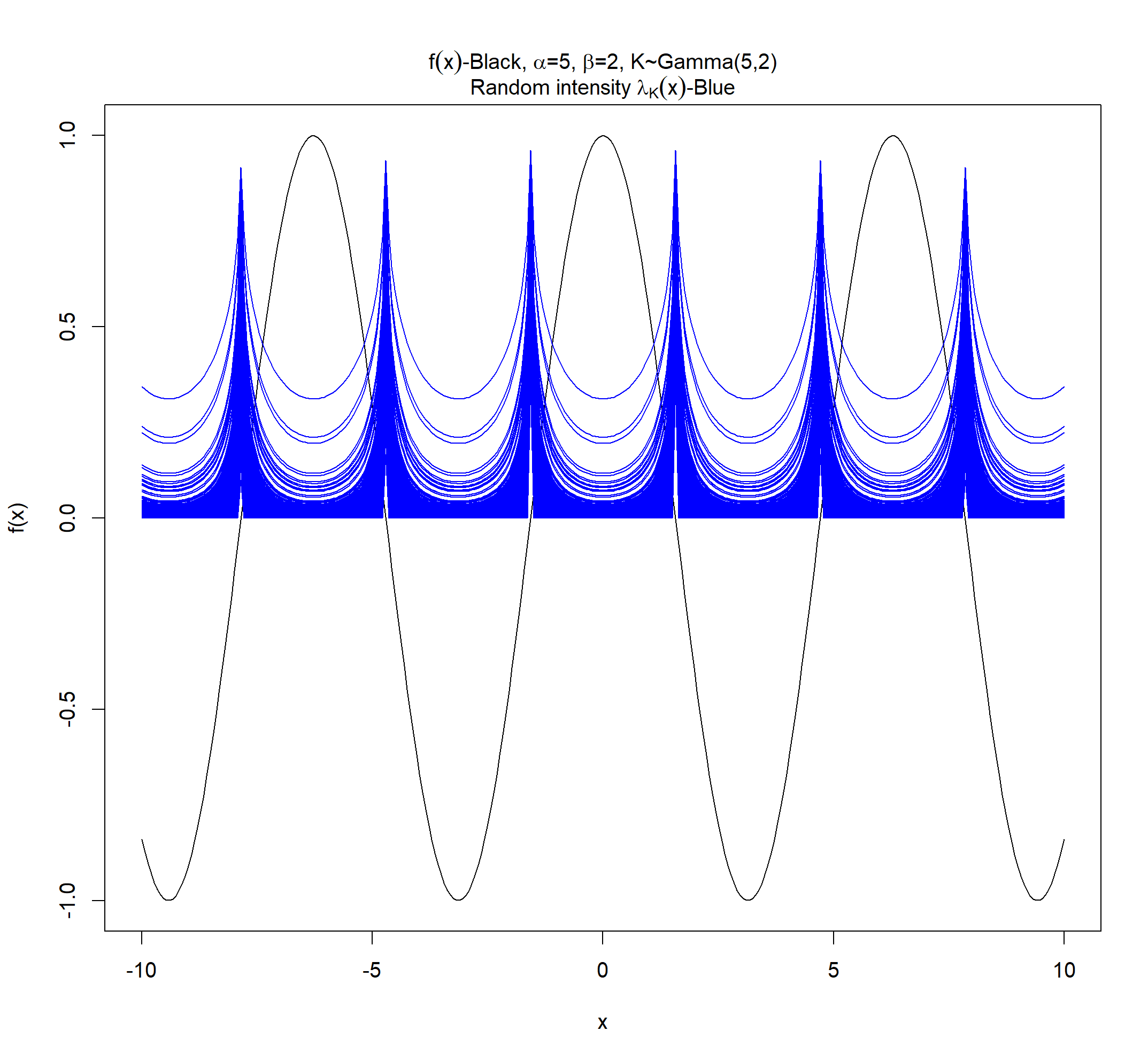} %
\includegraphics[width=0.495\textwidth]{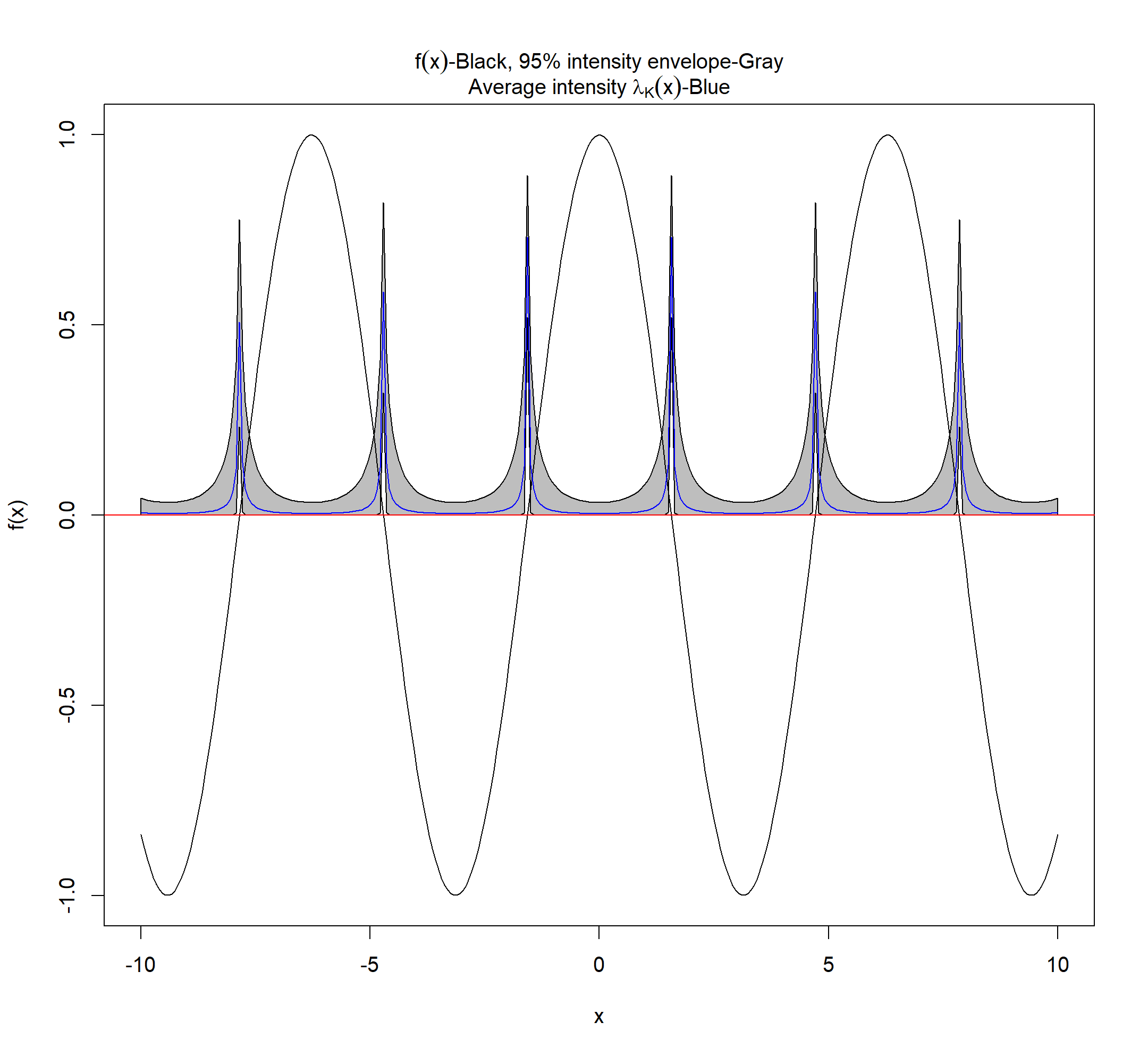}
\caption{The function $f(x)=cos(x)$, over the window $\mathcal{W}=[-10,10]$
(black color). We present 1000 realizations (left plot, blue color) of the
random intensity function $\protect\lambda _{K,Q}(\mathbf{x})$ $(Q=0.5),$
with source of randomness the parameter $K\thicksim Gamma(5,2).$ The plot on
the right presents 95\% envelopes of the random intensity (gray color), as
well as its mean (blue color).}
\label{CoxExamplesplot1}
\end{figure}

\section{Concluding Remarks}

We presented a novel method for finding function zeros utilizing point
process theory. The APPPZ is able to obtain all zeros of a real or complex
function, and in many cases in a single iteration.

Several modifications and applications of the presented methods will be
considered in the future, including a hierarchical Bayesian modeling
approach to help the APPPZ select appropriate values of $N$, $K$ and $Q,$
automatically, in order to increase its performance in terms of accuracy and
speed. In addition, the assumptions on the objective function will be
relaxed to help us investigate less well behaved functions, as well as
investigate the dependent zeros case using Cox random measures.

In this introductory paper we presented basic applications of the new method
in order to study its behavior. Specific applications to important
mathematical contexts, such as linear and nonlinear programming, machine
learning, optimal control and variational problems, are currently under
investigation. These results are of great interest and will be presented
elsewhere.

\section{Statements and Declarations}

The author has no financial or non-financial conflicts of interest that are
directly or indirectly related to this work submitted for publication.

\section*{Acknowledgements}

I am grateful to Professor Stamatis Dostoglou, Department of Mathematics,
University of Missouri, for his constructive comments and suggestions on an
earlier version of the manuscript.

\bibliographystyle{plainnat}
\bibliography{IPPzeroes}

\end{document}